# THE FIRST STELLAR ABUNDANCE MEASUREMENTS IN THE GALACTIC CENTER: THE M SUPERGIANT IRS 7

John S. Carr[1]

Naval Research Laboratory, Code 7217, Washington, DC 20375; carr@mriga.nrl.navy.mil

K. Sellgren[1]

Department of Astronomy, Ohio State University, Columbus, OH 43210;

sellgren@payne.mps.ohio-state.edu

AND

Suchitra C. Balachandran[1]

Department of Astronomy, University of Maryland, College Park, MD 20742;

suchitra@astro.umd.edu

ABSTRACT

The first measurement of the photospheric abundances in a star at the Galactic Center are presented. A detailed abundance analysis of the Galactic Center M2 supergiant IRS 7 was carried out using high-resolution near-infrared echelle spectra. The Fe abundance for IRS 7 was found to be close to solar, [Fe/H] = -0.02 ± 0.13, and nearly identical to the Fe abundances we obtained for the nearby M supergiants α Ori and VV Cep. Analysis of the first and second overtone lines of CO was used to derive an effective temperature of 3600 ± 230 K, a microturbulent velocity of 3.0 ± 0.3 km s$^{-1}$, and a carbon abundance log ε(C) = 7.78 ± 0.13, or [C/H] = -0.77. In addition, we find a high depletion of 0.74 ± 0.32 dex in O and an enhancement of 0.92 ± 0.18 dex in N. These abundances are consistent with the dredge-up of CNO-cycle products but require deep mixing in excess of that predicted by standard models for red supergiants. In light of our measured solar Fe abundance for IRS 7, we discuss other indicators of metallicity at the Galactic Center, the interpretation of low-resolution near-infrared spectra of late-type giants and supergiants, including the need for caution in using such spectra as measures of metallicity, and the evolution of massive young stars at the Galactic Center. We suggest the possibility that rapid stellar rotation is common for stars formed under conditions in the Galactic Center, and that extra internal mixing induced by high rotation rates, rather than evolution at high metallicity, is the explanation for many of the unusual properties of the hot emission-line stars in the Galactic Center.

Subject Headings: Galaxy: abundances --- Galaxy: center --- stars: abundances --- stars: late-type --- supergiants

1. INTRODUCTION

The Galactic Center is clearly an unique environment in the Galaxy. This statement certainly applies to the current population of high-mass stars observed in three dense and massive clusters that have formed during the last 10 Myr in the inner 50 pc. Of these three, the dense star cluster in the central parsec has been known and studied the longest. The central cluster is believed to be the site of a recent burst of star formation, as evidenced by the presence of red supergiants (Lebofsky, Rieke & Tokunaga 1982) and the discovery of more than a dozen emission-line stars which have been identified as highly-evolved massive stars, including Ofpe/WN9 and Wolf-Rayet (WR) stars (Allen, Hyland, & Hillier 1990; Krabbe et al. 1991, 1995; Libonate et al. 1995; Blum, DePoy, & Sellgren 1995; Tamblyn et al. 1996; see Maeder & Conti 1994 for a summary of massive star spectral types). Models for the starburst (Tamblyn & Rieke 1993; Krabbe et al. 1995) suggest that the stars formed during the last 3-8 Myr. The more recently discovered Quintuplet (Nagata et al. 1990; Okuda et al. 1990; Glass, Moneti, & Moorwood 1990; Moneti, Glass, & Moorwood 1994; Figer, McLean & Morris 1995, 1999; Cotera et al. 1996) and Arches clusters (Nagata et al. 1995; Cotera et al. 1996; Serabyn, Shupe & Figer 1998), at 25-30 pc projected distance from the center, contain similar collections of hot massive stars, including Ofpe/WN9 stars, WR stars of both WN (showing equilibrium CNO cycle products) and WC (He-burning products) subtypes, OB supergiants, and Luminous Blue Variables (LBVs). The massive stars observed in each of these clusters imply total cluster masses approaching that of a small globular cluster (Figer et al. 1999).

These young clusters represent what may be the most recent episodes of recurrent bursts of star formation in the Galactic Center. Indeed, it has been suggested that the centrally concentrated infrared cluster in the Galaxy's inner 100 pc is the result of sustained star formation over the lifetime of the Galaxy (Serabyn & Morris 1996), rather than the innermost part of the Galactic Bulge. If true, then stars in the Galactic Center may reflect a chemical evolution distinct from that in the Bulge or, perhaps, other parts of the Galaxy.

Information on the chemical abundances of this stellar population in our Galaxy is sorely lacking. Because the elemental abundances in stars record the nucleosynthetic history of the gas from which they formed, abundance studies of the stellar population in the Galactic Center may shed light on past star formation history, the source of gas falling into the center, and the potential role of Galactic Center outflows. For example, the unusual conditions under which stars must form within 100 pc of the Galactic Center may result in an initial mass function dominated by massive stars (Morris 1993) and a proportionately larger enrichment in oxygen and α-elements over time via SN II, depending upon how much enriched material is retained within the Galactic Center and incorporated into the next generation. The abundances influence our understanding of the Galactic Center in other ways. The overall metallicity affects comparisons of the luminosity function of the Galactic Center with other stellar populations. Knowledge of the abundances of CNO and s-process elements in some of the luminous cool Galactic Center stars could help to distinguish between younger, high-mass supergiants and older, intermediate-mass AGB stars, and hence constrain the times of the preceding star formation episodes. Knowledge of the stellar metallicity is also important for understanding the massive blue stars, since metallicity will affect the mass-loss rates, evolutionary duration of the post-main-sequence phase, and the relative ratio of WR sub-types.

In order to address such questions, we initiated a program to obtain quantitative measurements of chemical abundances for cool luminous stars in the Galactic Center using high-resolution spectra and standard abundance analyses. The very high extinction towards the Galactic Center forces abundance studies to be carried out in the infrared. The required sensitivity for these observations has only been possible since the advent of near-infrared echelle spectrographs. Our observations and analyses to date have concentrated on the brightest M stars in the central few parsecs (Carr, Sellgren & Balachandran 1996; Ramírez et al. 1999). In this paper we present results for the Galactic Center M2 supergiant IRS 7. In addition to being the brightest star in the Galactic Center, its evolutionary age, ~ 7 Myr, makes IRS 7 a member of the current starburst, and its metallicity should be representative of the more massive hot stars in the central cluster.

## 2. DATA

### 2.1 Infrared Echelle Data

High-resolution spectra for IRS 7 and VV Cep were obtained at the NASA Infrared Telescope Facility using CSHELL, the facility cryogenic infrared echelle spectrograph (Tokunaga et al. 1990). The detector for most of these observations was a 256x256 NICMOS-3 HgCdTe array, which provided a spectral coverage of ~ 1000 km s$^{-1}$; an additional spectrum was later obtained when the spectrograph was equipped with a 256x256 SBRC InSb array, which gives a smaller coverage of ~ 750 km s$^{-1}$. Because of the limited spectral coverage in a single grating setting, it was necessary to choose particular grating settings to cover spectral features of interest. The central wavelength for each grating setting, the spectral features utilized, and signal-to-noise ratios are given in Table 1. Second overtone CO lines in the H band and a high excitation 2-0 line in the K band were used for the effective temperature analysis. These lines were combined with moderately strong CO 2-0 lines near 23120 Å to determine the microturbulent velocity. The Fe I lines in the iron abundance analysis came from various K band spectra.

All of the spectra were obtained with a 0.5" slit, providing a nominal resolution of 40,000. The data were collected by alternately placing the star in one of two positions in the slit after each readout of the array. Subtraction of these A and B images removed detector dark current, residual images on the array, and any thermal or skyline emission. All of the A images were combined together and the spectrum was extracted; the same was done for the B images. Separate dispersion solutions were determined for the two beams before combining them into one spectrum. Either arc lamp emission spectra or telluric absorption lines in hot star spectra were used for the wavelength calibration, and radial velocity corrections based on the observed stellar features were applied to place the spectra on a rest wavelength scale.

For each spectrum of IRS 7, a hot star was observed at approximately the same airmass. The final hot star spectrum was divided into the target spectrum to correct for telluric absorption lines (when present) and for fringing in the spectrum. Fringing in the CSHELL spectra is caused by channeling in the circular variable filters used for order sorting. The observed fringes are about 2-3 periods over the wavelength coverage of the array with an amplitude (< 10 %) that depends upon the observed wavelength and other variables such as the star's position in the slit, seeing conditions, and source brightness. The degree of success in removing the fringing, as judged by comparison of the final A and B spectra, was variable. In cases where the shapes of the final A and B spectra differed significantly, equivalent widths were measured separately in each spectrum, and the difference was used to gauge the uncertainty due to channel fringing.

2.2 Archival Fourier-Transform Spectrograph Data

The near-infrared Arcturus atlas of Hinkle, Wallace & Livingston (1995) was used for the analysis of $\alpha$ Boo. The resolution of this atlas is 112,000 with a typical S/N ratio of 380. The line equivalent widths were measured separately in the summer and winter scans and averaged, except in cases where telluric features badly contaminated one of the two spectra. For $\alpha$ Ori we used archival KPNO FTS spectra (see Lambert et al. 1984) initially, but the final analysis of the K band data was based on the Wallace & Hinkle (1996) atlas of high-resolution spectra of cool stars, which are the same data corrected for telluric absorption, with an average resolution of 132,000 and S/N ratio of 380. KPNO FTS spectra were used for $\beta$ And, 30 Her, and HR 6702 (see Smith & Lambert 1985). The respective S/N ratios are 160, 100 and 100, and the resolution is 90,000. These spectra have not been corrected for telluric lines, rendering some stellar features unusable.

3. ABUNDANCE ANALYSIS

3.1 Approach

One of the difficulties that result from the very high extinction towards the Galactic Center is the determination of the stellar effective temperature to be used in the abundance analysis. Photometry cannot be used to establish the temperature because the observed stellar colors are completely dominated by reddening. An alternative is to derive a spectroscopic temperature from a fine analysis of lines in the stellar spectrum. In the optical, neutral lines of Fe or Ti are often used to determine $T_{eff}$ in dwarfs. In the infrared, neutral metal lines tend to arise from too narrow a range of excitation potential to provide good leverage in measuring $T_{eff}$; in addition, Fe I and other metals may be subject to non-LTE effects in giants and supergiants that would produce systematic trends with excitation potential (Ruland et al. 1980; Takeda 1992). Molecules are a good alternative for late-type giants since departures from LTE are expected to be small and there are a large number of ro-vibrational transitions. After investigating the possible choices, we decided that the CO molecule provided the greatest promise (see also Lambert et al. 1984), in terms of the available range in excitation potential and line strength, for deriving both the effective temperature and microturbulent velocity. Because most of the first overtone CO lines are strong in M giants, the temperature derivation must rely largely on weaker second overtone CO lines. Unfortunately, the second overtone occurs in the H band, where line blending is severe and Galactic Center stars are fainter due to the large reddening.

We analyzed a number of other late-type giants and supergiants in order to check the accuracy of our effective temperature determination and to account for any systematic effects in the abundances. The comparison stars, covering a range in gravity and temperature, are $\alpha$ Boo (K1 III), $\beta$ And (M0 III), $\alpha$ Ori (M1-2 Iab-a), VV Cep (M2 I), 30 Her (M6 III), and HR 6702 (M5+ II). Analyses of the first four stars allowed us to check the determinations of effective temperature and microturbulence from the CO spectrum by comparing to values published in the literature. $\alpha$ Boo provides an important abundance reference point since it has been the subject of numerous previous detailed analyses. Because our Fe abundances were determined using solar empirical oscillator strengths (§ 3.5), systematic effects in the results for cool low-gravity stars are a particular concern. However, even in the presence of errors relative to the solar abundance, direct comparisons between IRS 7 and other disk M supergiants should give a differential abundance free of systematic effects. We also examined the effects on our results of using different model atmosphere grids.

### 3.2 Model Atmospheres

Systematic errors in abundances can be introduced by various inadequacies in the model atmospheres of M giants, such as the treatment of molecular opacities, neglect of sphericity, and departures from LTE. We investigated the sensitivity of our results to the choice of model atmosphere by using three different sets of models. Our final abundance results for the M giants and supergiants are based on the models of Plez (Plez, Brett & Nordlund 1992; Plez 1992). These are spherically symmetric, opacity sampled model atmospheres, using water opacities based on a statistical representation of the water spectrum (following Alexander, Augason & Johnson 1989). The Plez grid is also the only one which includes gravities as low as that of IRS 7 (§ 3.3). Results for $\alpha$ Ori, VV Cep, and $\beta$ And were compared using three sets of models: Plez, Brown et al. (1989, hereafter BJACS), and Kurucz (1993). The latter two are plane-parallel atmospheres. The Brown et al. treatment of the water opacity is similar to that of Plez. The opacity distribution functions for Kurucz atmospheres, however, do not include water opacity, but this omission may not be a major effect for the temperatures and gravities of the early M supergiants analyzed in this paper. Arcturus is a low metallicity star ([Fe/H] ~ -0.5) with relative overabundances of the light metals. We used the atmospheric model developed specifically for Arcturus by Peterson, Ore & Kurucz (1993), which adopts $T_{eff}$ = 4300, log g = 1.5, and the observed mix of elemental abundances. Analysis of the solar spectrum and derivation of empirical "solar gf-values" were done using the Kurucz solar model (1993).

### 3.3 Surface Gravities

The surface gravity of IRS 7 was determined from estimates of its luminosity, temperature and mass, using the relation log g = log M + 4 log $T_{eff}$ - log L - 10.61. Extinction is the single largest uncertainty in the derived luminosity and gravity. Previously published values for $A_K$ range from 2.7 (Becklin et al. 1978) to 4.1 (Rieke & Lebofsky 1985; Rieke, Rieke, & Paul 1989), due to the uncertainty in the extinction curve. Recent values for $A_K$/E(H-K) range from 1.44 (Martin & Whittet 1990) to 1.78 (Rieke & Lebofsky 1985). E(H-K) = 2.3 ± 0.1 for IRS 7 (Becklin et al. 1978; Rieke & Lebofsky 1985), which implies $A_K$ = 3.3 to 4.1. We therefore adopted $A_K$ = 3.7 ± 0.4. The observed magnitude of K = 6.7 and a distance to the Galactic Center of 8.0 ± 0.5 kpc (Reid 1993) give an absolute magnitude of $M_K$ = –11.5 ± 0.4. To convert to bolometric magnitude, we used the bolometric corrections of Bessel & Wood (1984) with colors from Lee (1970). Adopting a spectral type of M1.5 with a range of M0 to M3, gives BC(K) = 2.70, $M_{bol}$ = -8.8 ±0.4 and log L = 5.4 ± 0.2. We used $T_{eff}$ = 3500 ± 250 K to estimate log g. The mass was estimated from the position of IRS 7 in the HR diagram in comparison to theoretical evolutionary tracks with solar metallicity (a factor of two change in metallicity does not alter the result). The models of Schaller et al. (1992), Bressan et al. (1993) and Meynet et al. (1994) indicate an initial mass for IRS 7 of 20 to 25 $M_\odot$ and a current mass in the range of 14 to 20 $M_\odot$. Adopting 17 ± 3 $M_\odot$ for the current mass, and including the uncertainties in the extinction, mass, distance, $T_{eff}$ and bolometric correction, leads to log g = –0.6 ± 0.2. The lowest surface gravity models of Plez with log g = -0.5 were used in the spectral analysis of IRS 7.

Values of log g for the other stars were taken from previous studies in the literature. We adopted log g = 0.0 for α Ori (Lambert et al. 1984), 0.0 for VV Cep (Luck 1982). Gravities for the M giants were taken from Smith & Lambert (1985): log g = 1.5 for β And, 0.2 for 30 Her, and 0.7 for HR 6702. The Arcturus model of Peterson et al. (1993) used log g = 1.5.

3.4 Effective Temperature, Microturbulence, and Carbon Abundance

The effective temperature $T_{eff}$ and microturbulent velocity ξ for IRS 7 and comparison stars (except for 30 Her and HR 6702) were derived from the analysis of first and second overtone CO lines. This also provided the C abundance since CO is the dominant form of C and the CO strength is independent of the O abundance in oxygen rich stars. Weak lines that are insensitive to microturbulence and cover a large range in excitation potential are desirable for the determination of $T_{eff}$. Since most of the first overtone lines are too strong for this purpose, measurements of the second overtone lines are required. Unfortunately, the severe line blending in the H band and the large macroturbulent broadening in M supergiants greatly limits the number of useful CO lines.

In order to choose lines which are largely unblended, synthetic spectra were calculated for the wavelength regions of interest. All of the spectral syntheses and abundance analyses were carried out with an updated version of the program MOOG (Sneden 1973). Basic data for the CO lines (wavelengths, energy levels and gf-values) were taken from the tables of Goorvitch (1994). Fe I line positions and energy levels were taken from Nave et al. (1994). Oscillator strengths for OH lines were obtained from Black (1995; see Balachandran & Carney 1996 for details) and wavelengths from Abrams et al. (1994). The Kurucz (1993) line lists were used for CN and other atomic lines.

Synthetic solar spectra were generated using the Kurucz (1993) solar model and compared to electronic versions of the disk-center infrared solar spectrum of Livingston & Wallace (1991). The damping constants for all atomic lines were initially set to twice that of the Unsöld approximation for van der Waals broadening; this value is appropriate for our Fe I lines (§ 3.5) but is not critical for the purpose of evaluating line blending. The oscillator strengths of the atomic lines and damping constants were adjusted, if necessary, to match the solar spectrum. Any spurious lines in the Kurucz list or lines that would be too weak to be observable in cool stars were deleted from the list, and unidentified lines in the solar spectrum were noted. Next, synthetic spectra were generated for Arcturus, using the model atmosphere and stellar abundances given in Peterson et al. (1993), and compared to the Arcturus infrared atlas (Hinkle et al. 1995). In general, the linelist with solar gf-values provide a good match to Arcturus; oscillator strengths were set for a small number of low excitation lines not observable in the solar spectrum, and additional unidentified lines were noted. Synthetic spectra with this final linelist were then generated for the nominal stellar parameters and CNO abundances of $\beta$ And (Smith & Lambert 1985) and $\alpha$ Ori (Lambert et al. 1984) using the Kurucz grid. The spectra were convolved with a macroturbulent broadening function (Gray 1992) to match the line widths in the FTS archival spectra (15 km s$^{-1}$ for α Ori and 6 km s$^{-1}$ for β And), and synthetic spectra were produced for both the FTS and CSHELL instrumental broadening.

The complete synthetic spectrum, that calculated with only the lines of interest (CO or Fe I), and the observed spectrum were compared to determine which lines were unblended or could be easily deblended for equivalent width measurement. As one progresses to stars with lower temperature and gravity, the line blending becomes more severe, mainly due to larger macroturbulent broadening and stronger CN lines. Few lines are truly unblended in the M supergiants. In addition, the limited wavelength coverage and lower resolution of the CSHELL spectra further limit the number of useful lines; for example, the number of CO lines available for the temperature analysis of IRS 7 is fewer than desirable.

Table 2 lists the CO lines used in our analysis, their wavelengths, energy levels and gf-values, and the measured equivalent widths for the giants α Boo and β And. Equivalent widths for the M supergiants (α Ori, VV Cep, and IRS 7) are given in Table 3. The uncertainties also include errors in the placement of the continuum level and, for the CSHELL data, any differences between the A and B spectra. The effective temperature was determined by requiring the derived carbon abundance, log ε(C) (log of the C abundance on a scale where the abundance of H is 12.0), to be independent of the lower excitation potential, χ, i.e., a slope of zero in a plot of log ε(C) vs. χ. The stronger 2-0 R(79) to R(83) lines and lines stronger than 300 mÅ were excluded from the temperature analysis. The microturbulence was adjusted until a slope of zero was obtained in a plot of log ε(C) vs. log reduced equivalent width, using weak to strong lines of a similar excitation potential. Examples are shown in Figure 1, and the results for each star are discussed next.

3.4.1 Results for Individual Stars

α Boo: A large number of CO lines can be measured from the Hinkle et al. atlas for α Boo. We measured 12 clean lines in the 16300 Å region, 3 higher excitation (~ 2 eV) lines near 23300 Å, and 5 stronger 2-0 lines for use in determining the microturbulence. We used the Peterson et al.

(1993) model atmosphere. From the 10 lines with excitation potential near 1.6 eV, we derived $\xi = 1.72 \pm 0.13$ km s$^{-1}$ (Fig. 1) Using this microturbulence, the 15 CO lines with $W_\lambda \leq 100$ mÅ give $\log \varepsilon(C) = 8.03 \pm 0.04$ (standard deviation). These values are in good agreement with the carbon abundance (8.06) and microturbulence (1.7 km s$^{-1}$) obtained by Peterson et al. from analysis of the optical spectrum. We did not independently determine the effective temperature; the small slope in $\log \varepsilon(C)$ vs. excitation potential (Fig. 1) would require a slightly higher (~ 80 K) temperature but is consistent within one $\sigma$ of their $T_{eff} = 4300$ K model.

β And: A similar set of CO lines was measured for β And from the KPNO FTS spectrum: 9 lines in the 16300 Å region, 3 strong 2-0 lines near 23110 Å, and 3 higher excitation 2-0 lines (fewer lines were used than for α Boo due to blending or telluric contamination). The microturbulence was determined from 7 weak to strong lines of about 1.6 eV, and the $T_{eff}$ and the C abundance were determined from 12 weak to moderate strength lines ($W_\lambda \leq 170$ mÅ). We obtained $T_{eff} = 3800 \pm 90$ K, $\xi = 2.16 \pm 0.20$ km s$^{-1}$, and $\log \varepsilon(C) = 8.31 \pm 0.05$ with the Plez models (log g = 1.5 and 2 $M_\odot$). The error in the abundance includes the uncertainty in $T_{eff}$ and $\xi$ and the standard error of the mean from the individual lines. Our effective temperature derived from CO lines is the same as that based on the V-K color (Smith & Lambert 1985), and our microturbulence agrees with the value determined from optical Fe I and Ti I lines by Smith & Lambert (1985), indicating that the microturbulence determined from CO is appropriate for the Fe I lines. However, our C abundance is smaller than their quoted value of 8.53. Upon analyzing their CO equivalent widths, we find that a major part of the difference is their inclusion of stronger CO lines, which produces a very strong trend of C abundance with line strength; their list also includes some lines we identify as blends or slight blends. Restricting the analysis to lines in common, and using the Plez models, we obtain $\log \varepsilon(C) = 8.36$ from their equivalent widths, in reasonable agreement with our result. The small difference is due to a systematic difference in the measured equivalent widths.

α Ori: Fewer clean CO lines are available from the α Ori spectrum, and coupled with greater uncertainties in equivalent widths and continuum placement, this results in larger uncertainties in the derived parameters. We chose a set of CO lines comparable to that available from the CSHELL spectra for IRS 7 and VV Cep (see Table 3). Figure 1 shows log ε(C) vs. χ and log ε(C) vs. log ($W_\lambda/\lambda$) for the best values of $T_{eff}$ = 3540 ± 260 K and ξ = 3.23 ± 0.15 km s$^{-1}$. The C abundance is 8.27 ± 0.16, where the error includes an assumed uncertainty in log g of 0.3. Our effective temperature is in agreement with the determination of 3520 ± 85 K by Dyck et al. (1992) from interferometer measurements at 2.2 microns. Lambert et al. (1984) adopted a higher $T_{eff}$ of 3800 K, largely based on the infrared flux method, for their CNO abundance analysis of α Ori. Their combined ($^{12}$C + $^{13}$C) abundance was 8.41. If we adjust their result for our lower temperature and their adopted $^{12}$C/$^{13}$C ratio of 6, we obtain log ε(C) = 8.23, consistent with our $^{12}$C abundance.

VV Cep: A similar analysis for VV Cep, using 7 CO lines measured from the CSHELL data, gave $T_{eff}$ = 3480 ± 250 K, ξ = 3.67 ± 0.20 km s$^{-1}$, and log ε(C) = 8.18 ± 0.12. A gravity of log g = 0.0 ± 0.3 was used. These results are similar to those obtained for α Ori. Our temperature is lower than the 3800 K estimated by Luck (1982) from optical Fe I lines, but closer to the temperature of ~ 3500 K expected for its spectral type (Di Benedetto 1993). Our microturbulence is also lower than the 5 km s$^{-1}$ used by Luck. The results from optical spectra are suspect, however, due to contamination from the B star companion (see § 3.5.2). No carbon abundance for VV Cep exists in the literature.

IRS 7: The four weakest CO lines in IRS 7 were used to derive the temperature, and four CO lines with similar excitation potential set the microturbulent velocity. The results with the Plez models for log g = -0.5 and 5 M$_\odot$ were $T_{eff}$ = 3470 ± 250 K, ξ = 3.30 ± 0.34 km s$^{-1}$, and log ε(C) = 7.73 ± 0.14. The effective temperature and microturbulence are similar to those found for α Ori and VV Cep, but the carbon abundance is much lower, [C/H] = -0.8 versus -0.3 and -0.4 for α Ori and VV Cep, respectively (where the solar abundance is 8.55).

The spectral differences between IRS 7 and the other two supergiants are well illustrated in Figure 2, which compares the three stars in a portion of the H band containing CO, OH, CN and atomic lines. The α Ori FTS spectrum has been smoothed to the lower resolution of the CSHELL data and overplotted on the spectra of IRS 7 (upper panel) and VV Cep (lower panel). There is a close similarity between the spectra of VV Cep and α Ori, with the atomic lines in VV Cep being somewhat weaker. IRS 7, however, shows a number of spectral differences compared to α Ori. As expected from the smaller C abundance, the CO lines in IRS 7 are noticeably weaker, in spite of the lower gravity which increase the CO strength. A much larger difference is seen in the OH lines, which are extremely weak in IRS 7, and in the CN lines which are significantly stronger (the latter is better seen in Figure 6; see also § 4.2).

Qualitatively, these differences illustrate a large depletion of C and O and enhancement of N in IRS 7 in comparison to α Ori and VV Cep. Such extreme changes in the CNO abundances may conceivably alter the atmospheric structure and thus the results of our analyses which were based on solar abundance models. We therefore repeated the CO analysis using a small set of atmospheres with altered CNO abundances (log $\varepsilon$(C) = 7.76, log $\varepsilon$(N) = 8.75, log $\varepsilon$(O) = 8.13), log g = -0.5, and 15 $M_\odot$, kindly provided by Plez (1997, private communication). The models with altered CNO abundances resulted in a somewhat higher temperature, smaller microturbulence, and a very slight increase in C abundance: $T_{eff}$ = 3600 ± 230 K, $\xi$ = 3.02 ± 0.30 km s$^{-1}$, and log $\varepsilon$(C) = 7.78 ± 0.13. We adopted these parameters for our IRS 7 analysis. Figure 1 shows the run of log $\varepsilon$(C) with excitation potential and line strength for this best fit model.

3.4.2 CNO Abundances for IRS 7 and α Ori

Abundances for O and N were obtained from synthesis of OH lines in the H band and CN lines in the K band, using the Plez models with altered CNO abundances for IRS 7. Figure 3(a) shows the spectrum of IRS 7 with a synthesis using the best fit O abundance, [O/H] = -0.74 ± 0.32. For comparison a synthesis is shown where the OH line strengths have been increased to match the O abundance we derived for α Ori, [O/H] = -0.21 ± 0.31. The error in the O abundance is dominated by the sensitivity of the OH lines to changes in $T_{eff}$. The CN lines used for the N abundance are shown in Figure 3(b) along with syntheses for the best N abundances for IRS 7, [N/H] = +0.92 ± 0.18, and for α Ori, [N/H] = +0.49 ± 0.19. In addition to the C and N abundances, the CN line strengths are dependent on the O abundance  The error in [N/H] is dominated by uncertainty in $T_{eff}$ indirectly through the dependence of [N/H] on the O and C abundances, which act in opposite directions and partially cancel out.

The sum of the CNO abundances in IRS 7 relative to the Sun is [(C+N+O) / H] = -0.09 ± 0.19 (solar values of log ε(C) = 8.55, log ε(O) = 8.87, and log ε(N) = 7.97 are from Grevesse et al. 1996). The uncertainty is mainly due to the error in [N/H], since N dominates the number sum of C, N and O for IRS 7. The sum of the CNO abundances in α Ori is [(C+N+O) / H] = -0.11 ± 0.24. These abundances are consistent with CNO processing of material, since within the uncertainties the sum of C, N and O, relative to Fe, is conserved. However, for IRS 7 the numbers imply very deep mixing into oxygen depleted regions of the ON-cycle in the star, which is not predicted for standard dredge-up by the convective envelope. For example, models for a 20 $M_\odot$ star, Z = 0.02, by Meynet et al. (1994) predict [$^{12}$C] = -0.35, [$^{14}$N] = +0.77, and [$^{16}$O] = -0.17 at core He exhaustion.

3.4.3 The Effect of Different Model Atmospheres

We investigated the dependence of the derived stellar parameters ($T_{eff}$, ξ) and C abundance on the family of model atmospheres by carrying out the same CO analysis for β And, α Ori, and VV Cep. The results of this comparison using the Kurucz, BJACS and Plez models are summarized in Table 4, where the derived $T_{eff}$, ξ, and C abundance are given for each star and model grid. We find that the derived effective temperature is nearly independent of the models used, and the C abundance has a range of no more than 0.1 dex. The BJACS models systematically give higher values for ξ and somewhat lower C abundances. When only the weakest CO lines are considered, the C abundances are essentially the same, and the different microturbulent velocities appear to compensate for differences in the models to reproduce the stronger CO lines. While only the Plez models were used for IRS 7, the consistency among the model atmosphere grids for α Ori and VV Cep suggests that the same result would be found if we had models with log g = -0.5 available from the other grids for comparison. We found that Plez atmospheres for 5 $M_\odot$ and 15 $M_\odot$ stars, at the same gravity, gave no difference in the derived parameters and C abundance for IRS 7.

3.5 Fe Abundance

Iron abundances were derived from Fe I lines in the K band. The wavelengths and lower energy levels were taken from the extensive list of Fe lines in Nave et al. (1994). Laboratory measurements of gf-values for infrared Fe I lines are rare, and we used semi-empirical solar gf-values. Fe I equivalent widths were measured from the Livingston & Wallace (1991) solar intensity spectrum. The gf-values were then derived using the Kurucz (1993) solar model with a microturbulent velocity of 1.0 km s$^{-1}$, a solar Fe abundance of 7.52, and van der Waal's broadening equal to twice the Unsöld approximation. This value for the damping constants seems appropriate for the high excitation potential (mostly $\chi > 4$ eV) of our Fe I lines (Holweger et al. 1991). As an additional check, synthetic solar spectra were generated and compared to the Fe I lines in the solar atlas, using macroturbulent velocities from Gray (1977). When necessary, slight adjustments were made to the gf-values and damping constants. Table 5 lists the Fe I lines, the measured solar equivalent widths and derived gf-values, and the equivalent widths of the same lines measured from the Arcturus atlas (Hinkle et al. 1995).

A careful check was made for line blending in the same manner as was done for the CO lines (§ 3.4). Synthetic spectra with and without Fe I lines were generated for the nominal parameters for α Boo, β And and α Ori in order to identify lines with minimum blending. CN lines are the main contamination in this spectral region, and a number of unidentified features also appear in the cooler stars. As expected, few lines in the M supergiant spectra are completely clean.

For α Boo and the M giants, our Fe abundances are based on measured equivalent widths (Tables 5 and 6). For the M supergiants, however, we deemed that synthesis of the spectrum in the region of each Fe I line provided a more reliable result, and we give instead [Fe/H] for each line (Table 7). As for CO, the uncertainty in the Fe I lines includes the uncertainty in continuum placement and, for VV Cep and IRS 7, differences in the A and B beams of the CSHELL spectra.

3.5.1 Results for Giants

The measured equivalent widths for 20 Fe I lines in the α Boo spectrum are given in Table 5. With the Peterson et al. (1993) model atmosphere and $\xi = 1.72$ km s$^{-1}$, the mean Fe abundance is [Fe/H] = -0.49 ± 0.02 (standard error of the mean). This result agrees very well with previous determinations of the Fe abundance in Arcturus, which average to [Fe/H] = -0.5 (Peterson et al. 1993; McWilliam & Rich 1994; Gratton & Sneden 1990; Branch, Bonnel & Tomkin 1978).

The Fe abundances for the M giants are based on the Plez model atmospheres and the equivalent widths in Table 6. We used the stellar parameters ($T_{eff}$, $\xi$) as determined from the CO analysis for β And. Using 12 Fe I lines that were not badly affected by telluric contamination or stellar blends, we obtained [Fe/H] = -0.03 ± 0.08. The error in the Fe abundances for β And and other stars includes the standard error of the mean for the individual lines and the errors due to uncertainties in $T_{eff}$, $\xi$, and log g. We adopted the stellar parameters for 30 Her and HR 6702 from Smith & Lambert (1985): $T_{eff}$ = 3250 K, log g = 0.2, $\xi$ = 2.7 km s$^{-1}$ for 30 Her; $T_{eff}$ = 3300 K, log g = 0.7, $\xi$ = 2.5 km s$^{-1}$ for HR 6702. The Fe abundances, based on 8 Fe I lines each, are [Fe/H] = 0.00 ± 0.11 for 30 Her and [Fe/H] = 0.16 ± 0.11 for HR 6702. These three stars were analyzed by Smith & Lambert using optical Fe I lines in the red, but they give Fe/H relative to α Tau: -0.10 ± 0.13 for β And, 0.07 ± 0.22 for 30 Her, and 0.02 ± 0.21 for HR 6702. The mean difference between our values of [Fe/H] and their abundances relative to α Tau is +0.05 ± 0.11.

### 3.5.2 Results for Supergiants

The Fe abundance for α Ori is based on the spectrum synthesis of 11 Fe I lines, giving [Fe/H] = -0.04 ± 0.08. For VV Cep, 4 lines were synthesized from the CSHELL spectra, yielding [Fe/H] = -0.06 ± 0.13. The result for VV Cep disagrees with the low abundance of [Fe/H] = -0.4 obtained by Luck (1982). Other metal abundances obtained by him for VV Cep are similarly low. These low values can likely be understood as due to the diluting effect from the flux of the B8 companion to VV Cep; the stars have similar fluxes at 4000 Å, but the B star does not contribute significantly to the infrared flux (see Hack et al. 1992).

Our Fe abundance derived for IRS 7 is based on 6 Fe I lines. Figure 4 shows the spectrum around the region of each of these lines for both IRS 7 (upper spectrum in each panel) and α Ori. For each star a synthetic spectrum using the best fit value of [Fe/H] is overplotted. For IRS 7, we also show synthetic spectra for [Fe/H] = +0.3, a value often adopted for the Galactic Center metallicity. The average abundance for IRS 7 is [Fe/H] = -0.02 ± 0.13, similar to both α Ori and VV Cep and not significantly different from the solar abundance. Table 8 shows the relative contributions to the total uncertainty in the final Fe abundance for IRS 7. The Fe abundance is relatively insensitive to the effective temperature, and the largest contributors to the uncertainty are ξ and the line-to-line scatter. The abundances for individual lines are further compared for IRS 7 and α Ori in Figure 5 by plotting [Fe/H] vs. line excitation potential. While the means are similar, the scatter for IRS 7 (and also for VV Cep) is larger than that for α Ori. We attribute the greater scatter to fringing in the CSHELL spectra and greater line blending in IRS 7.

### 3.5.3 Discussion of Fe Results

The Fe abundance results for all the stars are summarized in Table 9. In addition to the results obtained using the Plez models, results for β And, α Ori, and VV Cep obtained using both the Kurucz and BJACS model grids are shown. In each case, the Fe abundance was determined using the $T_{eff}$ and ξ determined from the CO lines for that grid of atmospheres. Table 9 shows that there are no obvious systematic differences in the Fe results between the models; all give essentially the same abundance for a given star when consistent stellar parameters are used.

The iron abundances could be affected by departures from LTE. This has been pointed out in G and K giants (Ruland et al. 1980; Takeda 1992), in which abundances derived from low-excitation lines, which form higher in the atmosphere, are systematically lower than those from the higher excitation lines. Our use of mostly high excitation ($> 4$ eV) Fe I lines, which form deeper in the atmosphere, should minimize this problem. Nevertheless, the Fe abundances *relative to the Sun* as given in Table 9 could well have systematic errors. However, *differential comparisons* between similar stars should eliminate any systematic effects due to non-LTE or errors in the model atmospheres. Hence, it is significant that we find no difference in Fe/H between IRS 7 and the local M supergiants $\alpha$ Ori and VV Cep, or the other giants we have analyzed.

One potential source of systematic abundance differences between IRS 7 and $\alpha$ Ori could be the position of the continuum. The CN lines in IRS 7 are significantly stronger than in the other two M supergiants, and a depression of the apparent continuum by numerous CN lines could result in artificially low equivalent widths for metal lines in IRS 7. To test this possibility, we produced synthetic spectra for large spectral regions using the CN linelist from Jørgensen & Larsson (1990). This linelist is meant to be complete in terms of total opacity and contains an order of magnitude more lines (mostly weak transitions) in our spectral regions than the Kurucz list we used for our routine synthesis calculations. After synthesizing spectral regions with the stellar parameters and CNO abundances previously used for IRS 7 and $\alpha$ Ori, we were able to conclude that the stronger CN lines in IRS 7 do not alter the chosen continuum in the high-resolution spectra. An alternative molecule for producing a veil of weak lines is $H_2O$, but the water vapor absorption in low-resolution spectra of IRS 7 is similar to that in other M supergiants (Sellgren et al. 1987) or perhaps slightly weaker in strength (see Figure 2 in Blum, Sellgren & DePoy 1996).

## 4. DISCUSSION

4.1 Metallicity at the Center of the Galaxy

We find that the Fe abundance of the Galactic Center M supergiant IRS 7 is essentially solar and, furthermore, very similar to other M giants and supergiants in the solar neighborhood that we have analyzed in parallel with IRS 7. Specifically, our quantitative result is [Fe/H] = -0.02 ± 0.13, which is only 0.02 dex higher than our abundance for the M supergiant $\alpha$ Ori. The solar iron abundance we derive for IRS 7 is in marked contrast to the common supposition of higher-than-solar metallicity at the Galactic Center.

Previous measurements of Galactic Center metallicity have come from gas-phase abundances based on far-infrared fine-structure emission lines from H II regions. The evidence for higher-than-solar nebular abundances is somewhat mixed. Aitken et al. (1974, 1976), Willner et al. (1979), and Shields & Ferland (1994) concluded that Ne/H was close to solar in Sgr A West, the H II region at the nucleus. Willner et al. (1979), Lester et al. (1981), and Shields & Ferland (1994) found Ar/H in Sgr A West to be roughly twice solar, but Lester et al. noted that the accuracy of the argon abundance was limited by uncertainty in the [Ar II] collision strength. The recently revised collision strengths for [Ar II] by Pelan & Berrington (1995), combined with the Lester et al. data, yield a near-solar argon abundance of Ar/H = $3.3 \times 10^{-6}$. Analyses of other H II regions within 100 pc of the Galactic Center (G359.98-0.08 and G0.095+0.012) have been carried out by Simpson et al. (1995) and Afflerbach, Churchwell & Werner (1997). Their determinations of N/H range from 2.0 to 4.1 times solar, O/H from 0.9 to 2.8 times solar, and S/H from 1.3 to 1.9 times solar.

There is not necessarily a direct conflict between the gas-phase abundances and the stellar abundances in the Galactic Center. The H II region observations include elements that either cannot be measured in the infrared stellar spectrum (Ne, Ar and S) or whose photospheric abundances are affected by stellar evolution (N and O). Our result for the sum of C, N and O [§ 3.4.2], however, is consistent with an original solar abundance of these elements that has been modified through the CNO cycle. Furthermore, the measurements of ionized gas near the Galactic

Center reflect the current gas-phase abundances, while the abundances of IRS 7 reflect the gas-phase abundances when this star was formed several million years ago. Enriched gas from supernovae or winds from massive hot stars may alter the gas-phase abundances on timescales of a few million years. Whether such enriched gas is incorporated into succeeding rounds of star formation may depend upon the efficiency of a Galactic Center wind in ejecting material from the center.

There are also potentially large uncertainties associated with the derivation of gas-phase abundances from the analysis of nebular lines. Analyses of the same infrared line data by Simpson et al. and Afflerbach et al. often yield abundances that differ by factors of 2–3, showing that different analysis methods can produce discrepant results. A major concern is the use of LTE model atmospheres for determination of ionization correction factors or detailed modeling of H II regions. Investigations of the ionization structure of H II regions using non-LTE atmospheres that include the effects of stellar winds (Sellmaier et al. 1996; Stasinska & Schaerer 1997) show that the predicted ratios of some ions are very different from ratios predicted with LTE atmospheres, particularly the ratio of $Ne^{++}$ to $O^{++}$, but also $N^{++}/O^{++}$ and $O^{++}/S^{++}$. Stasinska & Schaerer (1997), for example, find that the optical $N^+/O^+$ provides a direct and better indicator of the N/O ratio than the infrared $N^{++}/O^{++}$. Such differences are of even greater concern for the Galactic Center, because the ionizing continuum flux is dominated by hot post-main-sequence stars with extremely massive winds (Najarro et al. 1997). Finally, one may point out that the ionized gas structures at the Galactic Center (the "mini-spiral" of Sgr A West and the long arched filaments northeast of the nucleus) do not at all resemble the spherical classical H II regions of the models.

Another, albeit indirect, indication of metallicity at the Galactic Center are the properties of OH/IR stars in the inner 100-200 pc, such as their ($M_{bol}$, log P) relation and their wind expansion velocities. OH/IR stars are luminous, red, long-period variables (LPVs) on the AGB which trace an old to intermediate age stellar population. While the Galactic Center OH/IR stars are similar to those in the Galactic Bulge, they extend to higher luminosities and longer periods, suggesting that a younger population of LPVs exist in the Galactic Center (Blommaert et al. 1998; Wood, Habing & McGregor 1998).

Wood et al. (1998) have argued that the ($M_{bol}$, log P) relation for bluer, longer period Galactic Center LPVs suggests a lower luminosity, and therefore a higher metallicity, compared to LPVs in the LMC and Baade's Window. Our own analysis of their data (see Figure 11 in Wood et al. 1998) finds instead that the bluer Galactic Center LPVs show a ($M_{bol}$, log P) relationship very similar to the LPVs in Baade's window, with no evidence for lower luminosity.

The Galactic Center OH/IR stars show higher expansion velocities, on average, then samples from other stellar populations (Wood et al. 1998; Sjouwerman et al. 1998). The wind expansion velocities are thought to depend on stellar luminosity, mass-loss rate, and the dust-to-gas ratio (Habing, Tigon & Tielens 1994). If it is assumed that the dust-to-gas ratio in the outflow depends only on the metallicity (specifically, the Si abundance in oxygen-rich stars), then for similar luminosities and mass-loss rates, higher expansion velocities imply higher metallicity. The expansion velocities of the Galactic Center OH/IR stars are ~ 20 % larger (at log P ~ 2.8) then for local and Bulge Miras (Wood et al. 1998), implying a dust-to-gas ratio higher by a factor of 1.6 (or 0.20 dex). The metallicities of the Bulge and Galactic disk OH/IR stars are not known, but if they are similar to the mean of giants in the Bulge, [Fe/H] = -0.25 (McWilliam & Rich 1994), and solar neighborhood, [Fe/H] = -0.17 (McWilliam 1990), then a potential increase in dust-to-gas ratio of $\leq$ 0.2 dex is consistent with our finding of a solar metallicity for IRS 7.

4.2 Comparison to the Low-Resolution Spectroscopy

Sellgren et al. (1987) and Blum et al. (1996) found that the strengths of the Na and Ca features as observed in low-resolution (R = 600 - 3000) K-band spectra were stronger in IRS 7 and other Galactic Center giants than in field stars of similar spectral type. Larger equivalent widths in these Na and Ca features, compared to field stars, are also observed in Baade's window M giants at R = 1000 (Terndrup et al 1991; Blum et al. 1996). Terndrup et al. interpreted their results as evidence for a mean metallicity of +0.3 in the Bulge. However, high-resolution optical spectroscopy and abundance analysis of Baade's window K giants (McWilliam & Rich 1994) show abundances similar to the solar neighborhood, with a mean value of [Fe/H] = -0.25 and a range of -1.0 to +0.45. How should the Na and Ca features in low-resolution spectra of Galactic Center stars be interpreted, in the light of our solar Fe abundance for IRS 7?

We have compared three M supergiants of similar effective temperature but different luminosities to show that differences in gravity or luminosity cannot be the sole reason for the differences in Na line strengths measured at lower resolution. We smoothed the FTS spectrum of IRS 7 (Sellgren et al. 1987) and the high-resolution Wallace & Hinkle (1996) spectrum of $\alpha$ Ori to the same resolution (R=3000) as the spectrum of the M2Ia supergiant $\mu$ Cep from Kleinmann and Hall (1986). Our measured equivalent width of the Na feature is nearly identical in $\alpha$ Ori and $\mu$ Cep, 6.5 Å, even though the stellar luminosities are very different, log L = 5.6 and 4.75, respectively. The measured equivalent width of the Na feature at R = 3000 in IRS 7 is 9.3 Å, while its luminosity, log L = 5.4, is closer to that of $\mu$ Cep.

High-resolution spectra of late-type giants and supergiants (see Wallace & Hinkle 1996) show that the so-called Na and Ca features are a blend of several lines in addition to Na I or Ca I, including Sc I, Ti I, V I and CN. The relative contributions of individual lines depends on effective temperature, surface gravity, and the elemental abundances, including evolutionary changes in the CNO abundances which are reflected in the strength of the CN lines. In the coolest giants, Na I and Ca I are not even the principal contributors to the total equivalent width of their respective features (see Ramírez et al. 1997).

A comparison of high-resolution spectra of the Na region for IRS 7 and $\alpha$ Ori reveals the source of the differences in the equivalent widths measured at low resolution. Figure 6 shows our CSHELL spectrum for IRS 7 along with the FTS spectrum for $\alpha$ Ori (Wallace & Hinkle 1996), smoothed to the same resolution as the CSHELL data. The equivalent widths of the Na I lines are nearly identical in IRS 7 and $\alpha$ Ori. The most obvious difference between the two spectra are the stronger Sc I and V I lines in IRS 7. Stronger CN lines in IRS 7 account for most of the remaining spectral differences. Quantitatively, the relative contributions to the total equivalent width *difference* measured over the Na region (from 22047 to 22129 Å) in the high-resolution spectra are: Sc I 26 %, V I 12 %, and CN lines ≥ 50 %. In low-resolution spectra, the measured integrated equivalent width of the Na feature will differ due to a lower placement of the apparent continuum level, but our spectral syntheses show that the relative contributions of species to the total equivalent width does not change significantly. Hence, the single largest cause for the larger Na feature equivalent width in IRS 7 is CN, a direct result of the very different mix of CNO abundances, presumably due to evolutionary differences. The anomalously strong lines of Sc I and V I are a secondary contributor to the larger Na feature strength in IRS 7.

We find that strong atomic lines of Sc I, V I and Ti I are systematically stronger in IRS 7 compared to the field supergiants. We have carried out a careful wavelength by wavelength comparison of the smoothed (R=3000) K-band spectrum of IRS 7 (Sellgren et al. 1987) with the K-band spectrum of μ Cep (Kleinmann and Hall 1986). The majority of the prominent differences between the two spectra (with IRS 7 always having stronger absorption) coincide in wavelength with Sc I lines, followed by lines of V I, Ti I and CN blends. This is supported by other spectral regions covered by our CSHELL data, in which we find the Sc I lines at 22041 and 21819 Å, and the Ti I line at 21784 Å, to be far stronger in IRS 7 than in α Ori.

It would be incorrect to presume that the stronger Sc I, V I and Ti I atomic lines in IRS 7 represent abundance enhancements. All of these lines are saturated, low-excitation transitions (typical excitation potentials of 1.4, 2.5 and 1.8 eV for Sc I, V I, and Ti I, respectively), and are hence very sensitive to the outermost layers of the atmosphere where the structure and turbulent motions are likely to differ from the theoretical models. Both the Sc I and V I transitions have hyperfine structure (hfs) which produces a desaturating effect on the lines. We have synthesized the spectra in the Na I region after including the hyperfine structure for the Sc I 22058 and 22071 Å lines. The hfs constants for these transitions were taken from Zeiske et al. (1976), Ertmer & Hofer (1976), and Aboussäid et al. (1996). The relative strengths of the hyperfine components were calculated following Tolansky (1948), and the absolute gf-values were calibrated against Arcturus, adopting [Sc/H] = -0.19 (McWilliam & Rich 1994). Even after accounting for the hfs, an abundance of [Sc/H] = +0.9 is required to fit the strength of the Sc I lines in α Ori; the stronger lines in IRS 7 require [Sc/H] = +1.5. This result is similar to the inability to fit the strongest CO lines in M giants (e.g., Smith & Lambert 1990; Tsuji 1991). It is not surprising that abundances cannot be extracted from such lines, since their strengths depend on the unknown structure of the outermost, optically thin layers. Variables which may significantly effect line formation in the upper photosphere, such as chromospheric heating, temperature inhomogeneities, depth-dependent turbulence, and velocity fields (see Gustafsson & Jørgensen 1994) could well differ between IRS 7 and our comparison field stars.

The observed increase in the low-resolution Na and Ca indices for IRS 7, compared to field stars, can be understood by a combination of altered CNO abundances and greater strengths of saturated low-excitation transitions. Determining whether or not the large indices in other Galactic Center giants have the same cause will require further work. Currently, we lack the data for other stars to carry out an analysis of CNO abundances or line strengths in the Na region as for IRS 7. At this point we can only note that the Sc I lines in the CSHELL spectra we have obtained for IRS 19 and IRS 22 are also stronger than in field giants of similar spectral type.

4.3 Implications for Evolution of Massive Stars in Galactic Center

The M supergiant IRS 7 is believed to have formed as part of the starburst which occurred in the central cluster in the last 3-8 Myr. As such, the initial metallicity of IRS 7 should be representative of the other massive stars which are now observed as hot emission-line stars. If the solar Fe abundance of IRS 7 is representative of the overall metallicity (i.e., no selective elemental enrichment), then the evolution of the massive stars must be considered in the context of solar metallicity, rather than twice solar metallicity as is often assumed.

The infrared spectra of the majority of He I emission-line stars in the central cluster most closely resemble Ofpe/WN9 stars (Krabbe et al. 1995; Blum et al. 1995a; Tamblyn et al. 1996), which are a rare class believed to be transition objects between Of supergiants and late-type WN stars, or a hotter quiescent phase of LBVs (Langer et al. 1994; Nota et al. 1996; Crowther & Smith 1997). Quantitative modeling of some of the Galactic Center He I stars has been carried out by Najarro et al. (1994; 1997), and various of their properties have been compared to Ofpe/WN9 and WR stars by Schaerer (1999). The derived He abundances (helium mass fraction Y = 0.8 - 1.0) are larger than those in both Ofpe/WN9 stars (Y = 0.5 - 0.7) and in the few studied LBVs (Y ~ 0.6, Langer et al. 1994). The terminal wind velocities are also larger than Ofpe/WN9 stars in the LMC. In terms of both of these properties, the Galactic Center stars are rather similar to late-type WN stars (WN6-8), and they follow the same wind momentum relationship, though the typical mass-loss rates derived by Najarro et al. are higher than for WNL stars (Crowther & Smith 1997). The Galactic Center stars, however, have higher luminosities than WR stars (with a few exceptions: e.g., Crowther & Dessart 1998) and, perhaps more challenging to explain, significantly lower temperatures than either Ofpe/WN9 or WNL stars. Some of the properties of the Galactic Center emission-line stars could be understood in terms of evolution at high metallicity (Schaerer 1999). The presence of a number of late-type WC stars (WC9) in the central cluster (Blum, Sellgren & DePoy 1995; Krabbe et al. 1995) may be significant in this regard, since later WC types are associated with higher metallicity environments (Maeder & Conti 1994); however, in the case of a starburst, the fraction of various subtypes is also sensitive to the age and duration of the starburst (Schaerer & Vacca 1998). As pointed out by Schaerer (1999), however, many of the properties of the hot stars in the Galactic Center could also be explained by the action of extra mixing processes which produce effects similar to the larger mass-loss rates at high metallicity.

Rotation in massive stars can affect evolutionary tracks, surface abundances, and mass-loss rates (Maeder 1987; Langer 1992; Fliegner, Langer & Venn 1996; Langer & Heger 1998; Maeder 1998; Meynet 1998). Some recent calculations of evolution with rotationally induced mixing have been presented by Meynet (1998; see also Maeder 1998). These results follow the evolution until the star has just left the main sequence. Massive stars (20-40 $M_\odot$) with rapid rotation evolve to higher luminosities, attain higher surface He abundances, and reach cooler temperatures in the blue-supergiant phase. Enhanced mixing mimics the effects of increased mass-loss rates at high metallicity in pre-WR stages by making the formation of WR stars and WCL stars easier, with the appearance of H-burning and He-burning products on the surface at earlier stages of evolution.

We suggest that extra mixing induced by rapid rotation may indeed be the fundamental difference between the evolution of massive stars in the Galactic Center and those elsewhere in the Galaxy and in the LMC. The conditions for star formation in the Galactic Center are extreme and very different from star forming regions in the Galactic disk and include strong tidal forces, strong magnetic fields, and large gas temperatures and velocity dispersions. The angular momentum history of stars formed under these conditions may be different than stars forming elsewhere in the Galaxy. The formation of stars with preferentially rapid rotation would lead to a subsequent evolution with enhanced internal mixing and mass-loss rates. The derived CNO abundances for IRS 7 are clear evidence that at least one Galactic Center star has undergone extensive mixing, far beyond that expected for dredge-up by the convective envelope (§ 3.4.2). The high depletion of oxygen, [O/H] = -0.7 ± 0.3, indicates rather deep mixing into layers of the ON cycle. While this oxygen depletion is far greater than in either $\alpha$ Ori or VV Cep, it is difficult to say how unusual the CNO abundances for IRS 7 are because there are currently no measured CNO abundances for other M supergiants.

5. CONCLUSIONS

We have made the first direct determination of photospheric abundances for a star in the Galactic Center, based on standard abundance analysis of high-resolution, near-infrared spectra of the Galactic Center M supergiant IRS 7. We find a Fe abundance for IRS 7 that is essentially solar, [Fe/H] = -0.02 ± 0.13. Because the gf-values used for the Fe I lines were determined semi-empirically from the solar spectrum, our Fe abundances with respect to the solar value could include undetermined systematic errors; for this reason, comparison of these results to those obtained for similar field stars using the same analysis is important. When this is done, we find no difference in the Fe abundance between IRS 7 and the two solar neighborhood M supergiants $\alpha$ Ori and VV Cep, for which [Fe/H] = -0.04 ± 0.08 and -0.06 ± 0.13 were derived, respectively.

The C abundance, stellar effective temperature, and microturbulent velocity in IRS 7 were determined from the analysis of first and second overtone CO lines, which yielded log ε(C) = 7.78 ± 0.13, $T_{eff}$ = 3600 ± 230 K, and ξ = 3.02 ± 0.30 km s$^{-1}$. The low C abundance, [C/H] = -0.77, and our values for the oxygen and nitrogen abundances, [O/H] = -0.74 ± 0.32 and [N/H] = +0.92 ± 0.18, indicate that IRS 7 has undergone extensive dredge-up of CNO-cycle products that is in excess of the values predicted by standard models or observed in the supergiant α Ori. This suggests that some extra mixing process is required to explain the CNO abundances in IRS 7.

The extreme CNO abundances in IRS 7 result in stronger CN lines than are observed in the other M supergiants in our sample. The increased CN line strengths account for about half of the larger equivalent width of the low-resolution 22080 Å "Na feature" in IRS 7 compared to field stars. The remainder of the observed differences in the low-resolution "Na" feature is due to stronger Sc I and V I lines in IRS 7. These saturated, low-excitation features depend more on the poorly understood conditions in the outermost portions of the stellar atmosphere than on the elemental abundances, and hence the origin of these stronger lines in IRS 7 is not currently understood. Because the integrated equivalent widths of the "Na" and "Ca" features at low spectral resolution measure blends of unresolved saturated transitions (whose line formation is not properly modeled) and can depend on the degree of dredge-up of CNO products, we caution against their use as measures of "metallicity" in late-type giants and supergiants.

If the solar Fe abundance and the solar sum of C, N, and O abundances derived in this paper for IRS 7 represent the overall metallicity of the other massive stars, then the evolution of the massive stars in the current starburst in the central few parsecs should be considered in the context of solar metallicity, rather than twice solar or higher. The massive hot emission-line stars in the Galactic Center have higher He abundances, wind velocities, mass-loss rates, and luminosities, and lower temperatures, than analogous spectral types elsewhere in the Galaxy and the LMC. Given the extreme and unusual conditions under which star formation in the Galactic Center must occur, and the evidence that at least one Galactic Center star (IRS 7) has undergone extra mixing, we suggest that the evolution and observed properties of the massive Galactic Center stars are the result of extra mixing processes, induced, perhaps, by systematically higher stellar rotation rates for stars formed near the Galactic Center. Quantitative predictions of evolutionary models of massive post-main-sequence stars, which include rotationally-induced mixing, are required to test this idea.

The solar iron abundance for IRS 7 contradicts a common assumption of super-solar metallicity in the Galactic Center, a position that is suggested by some estimates of gas-phase abundances in the Galactic Center and a general trend of increasing abundances with decreasing radius in the Galactic disk. Whether there is a conflict between the stellar abundances and the abundance measurements for the ionized gas is unclear because of the potentially large uncertainties in the nebular abundances, the different elements that are measured by each technique, and the possibility that the ionized gas measures a short-lived phase of enriched gas. The photospheric abundances of Galactic Center stars are directly relevant to questions of the source of material for star formation at the Galactic Center and the extent to which enriched material from previous star formation episodes becomes incorporated over time into succeeding generations of stars.

The results for IRS 7 presented in this paper provide an estimate of the metallicity of the most recent burst of star formation in the central parsec of the Galactic Center. Investigating the chemical evolution and history of star formation in the Galactic Center, on the other hand, will require abundance analysis of a significant sample of stars over the inner 100 pc. It is also important to extend abundance analyses to other elements (e.g., the $\alpha$-elements) and to compare the abundance patterns to stellar populations in the Galactic Bulge and disk. Cool stars in the Galactic Center other than IRS 7 are less luminous and represent an older population. With the spectrographs in operation today, approximately 10 late-type stars are accessible to high-resolution spectral analysis, though not to the same detail as for IRS 7. These stars are likely to have ages in the range of 20 to 100 Myr (depending upon whether some of these are supergiants or massive AGB stars) and will sample older, but still relatively recent, episodes of star formation, perhaps overlapping in age with the most luminous OH/IR sources in the Galactic Center. Such a study is currently in progress (Ramírez et al. 1999). In the longer term, more sensitive spectrographs will permit abundance measurements of older, lower-mass stars on the red giant branch and allow us to determine whether or not the Galactic Center has followed a distinct star formation and nucleosynthetic history.

ACKNOWLEDGMENTS

This research was supported by NSF grants AST-9115236, AST-9618335 and AST-9619230, the Ohio State Columbus Fellowship, the Alfred P. Sloan Foundation, and the Office of Naval Research. We would like to thank the staff and telescope operators at the IRTF for their help and support. We also thank B. Plez for computing model atmospheres for us, K. Hinkle for providing archival FTS spectra and the Arcturus atlas data, V. Smith for FTS spectra of M giants, R. Kurucz for copies of his CD-ROMs, U. Jørgensen for providing electronic CN linelists, D. Terndrup and D. Schaerer for helpful discussions, and A. Krishnamurthi for assistance with observations.

FIGURE CAPTIONS

Figure 1. (a) The carbon abundance derived from CO lines versus lower excitation potential (left-hand panels) for the $T_{eff}$ values adopted for α Boo, α Ori and IRS 7. For α Ori and IRS 7, the plotted points are the abundances for the value of $T_{eff}$ that yields a zero slope (dotted lines). The dashed lines show fits that would result from using ± 1 σ values of $T_{eff}$. Plez (1992) model atmospheres are used for α Ori and IRS 7. For α Boo, the fit (dotted line) corresponds to the Peterson et al. (1993) model with $T_{eff}$ = 4300 K. (b) The carbon abundance versus log of the reduced equivalent width (right-hand panels). The plotted points are the abundances for the value of microturbulent velocity that yields a slope of zero (dotted lines). For α Ori and IRS 7, the dashed lines show the fits that would result from using ± 1 σ values of ξ.

Figure 2. Comparison of spectral features in a portion of the H band for three M supergiants of similar spectral type: α Ori, IRS 7 and VV Cep. The FTS spectrum of α Ori (smooth line), smoothed to the resolution of the CSHELL data, is compared to IRS 7 (histogram) in the upper panel and to VV Cep (histogram) in the lower panel. IRS 7 shows weaker CO and OH features than α Ori, but similar Fe I and Ni I strengths, while the spectra of VV Cep and α Ori are very similar.

Figure 3. (a) Spectrum of IRS 7 (histogram) in the region of the OH lines used to estimate the O abundance. The smooth line is the synthesis with the best O abundance for IRS 7, [O/H] = -0.74. The dotted line is a synthesis with the OH lines adjusted to the O abundance of α Ori, [O/H] = -0.21. The wavelengths of OH lines are marked by vertical lines. (b) Spectrum of IRS 7 in the region of the CN lines used to estimate the N abundance. The smooth line is the synthesis with the best N abundance for IRS 7, [N/H] = +0.92. The dotted line shows a synthesis using the N abundance of α Ori, [N/H] = +0.49. The wavelengths of CN lines are marked by vertical lines.

Figure 4. Observed and synthetic spectra in the region of Fe I lines in IRS 7 and α Ori. In each panel, the upper spectrum is IRS 7 and the lower one is α Ori. The histograms are the observed

spectra, and the solid smooth lines are the synthetic spectra for the best-fit value of [Fe/H] which is indicated. For IRS 7, the dotted lines are synthetic spectra with [Fe/H] = +0.3. Only one of the two IRS 7 spectra obtained for the 22386 and 22391 Å Fe I lines are shown.

Figure 5. The derived Fe abundance, [Fe/H], versus the lower excitation potential for the Fe I lines used in IRS 7 and α Ori. The mean value for IRS 7 is indicated by the dotted horizontal line.

Figure 6. Spectra in the region of the Na I doublet are compared for IRS 7 and α Ori. The histogram is the CSHELL spectrum for IRS 7, and the smooth line is the FTS spectrum for α Ori, smoothed to the resolution of the IRS 7 data. The Sc I, V I and CN lines are stronger in IRS 7. One set of CN lines is indicated, but nearly all the other unmarked spectral features are CN.

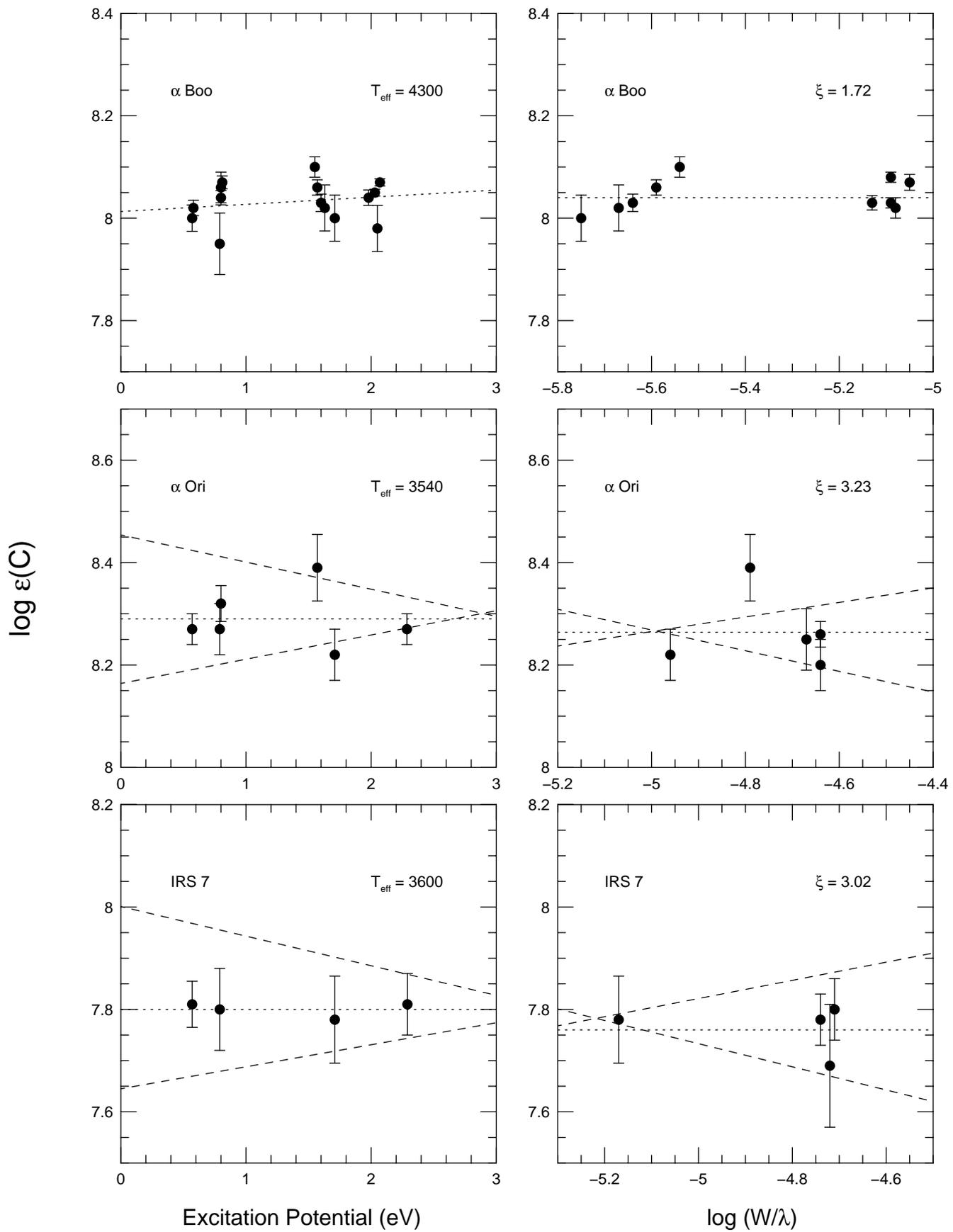

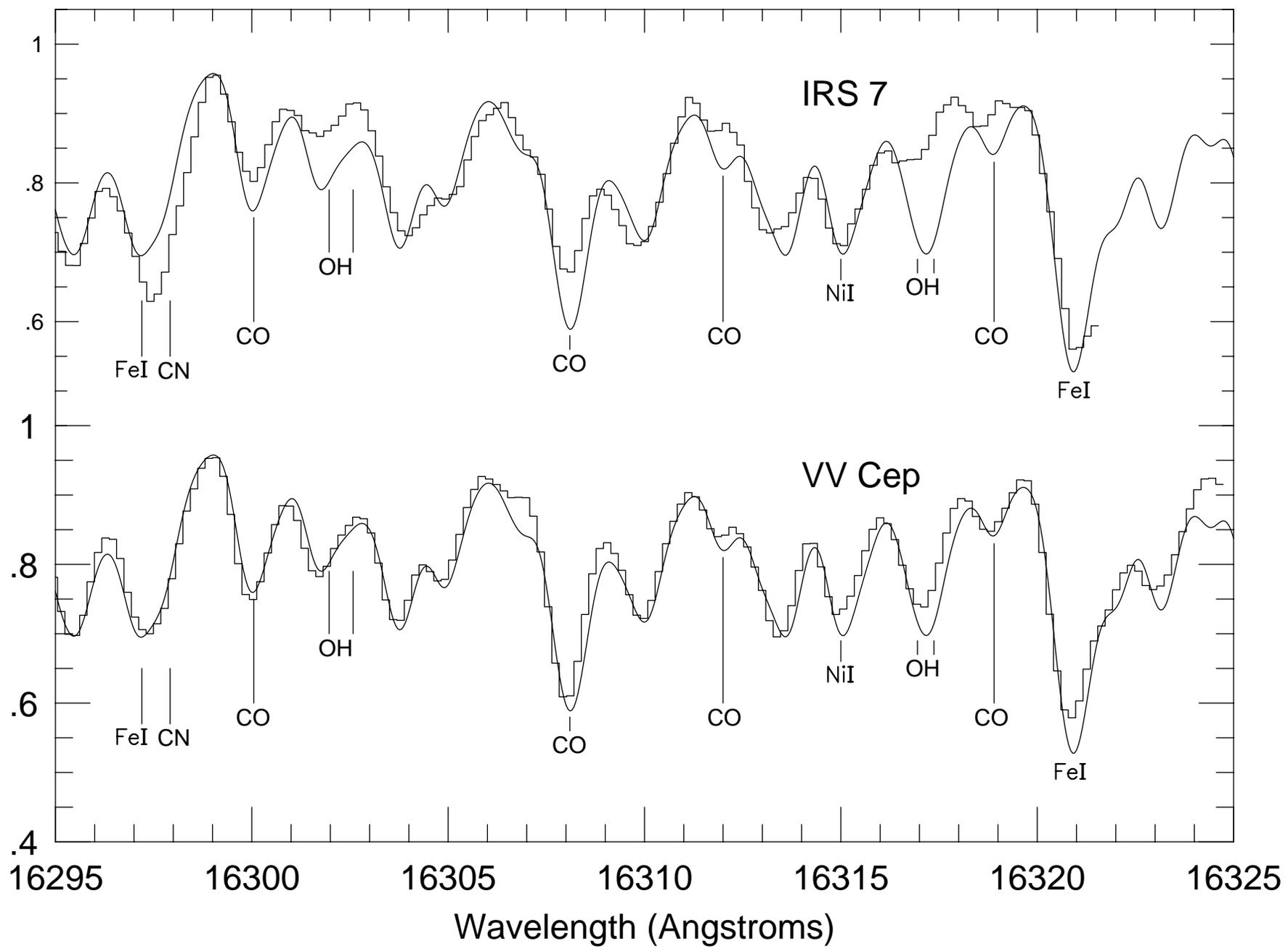

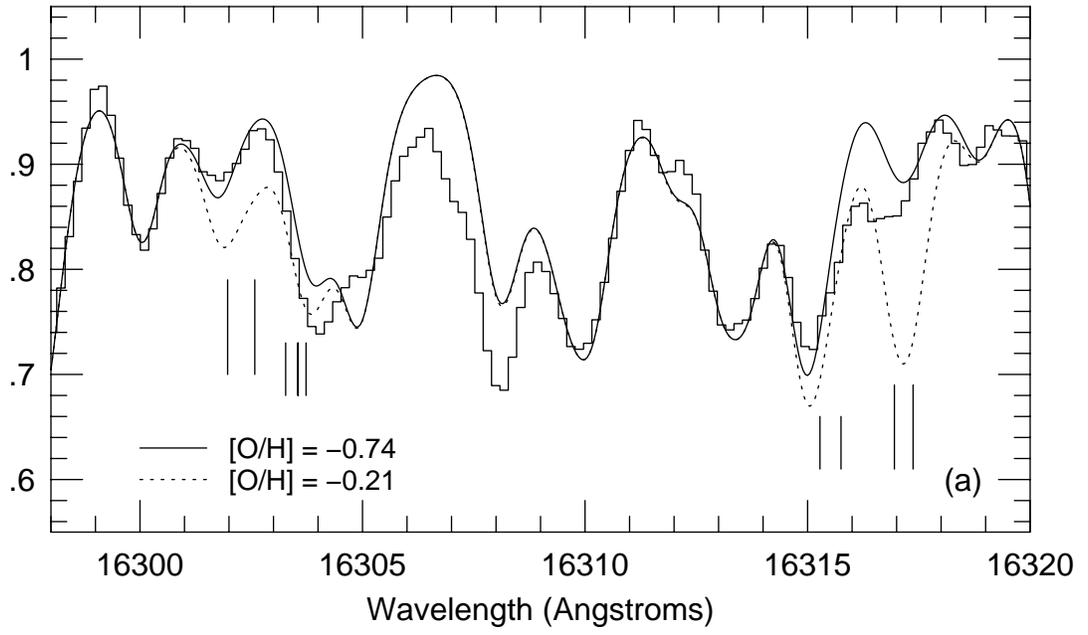
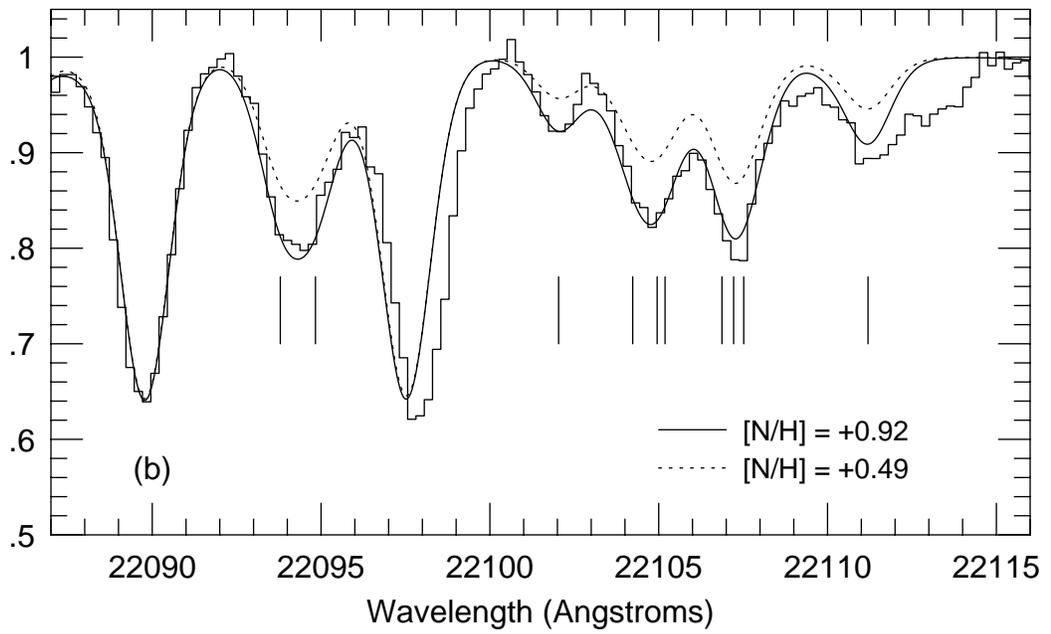

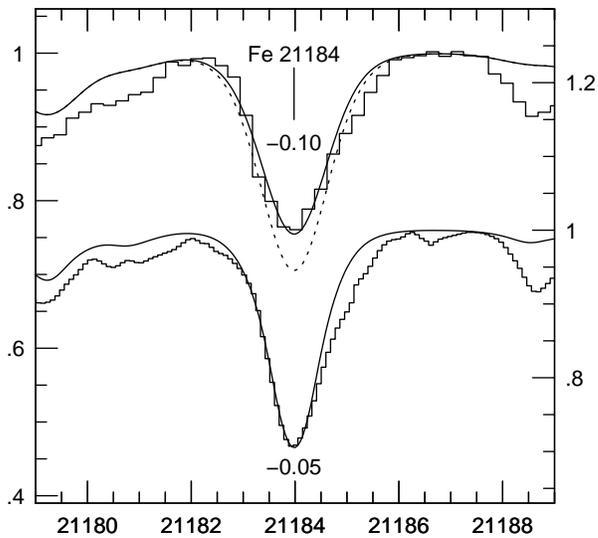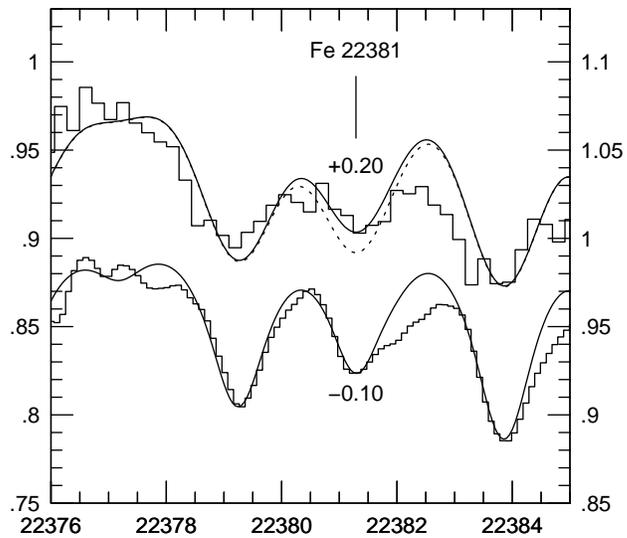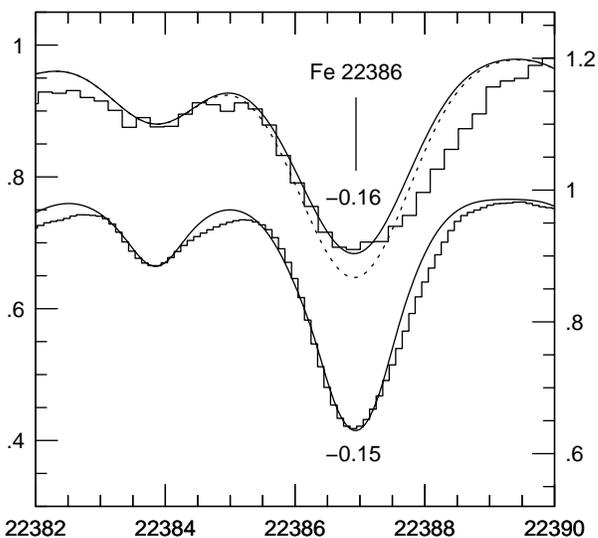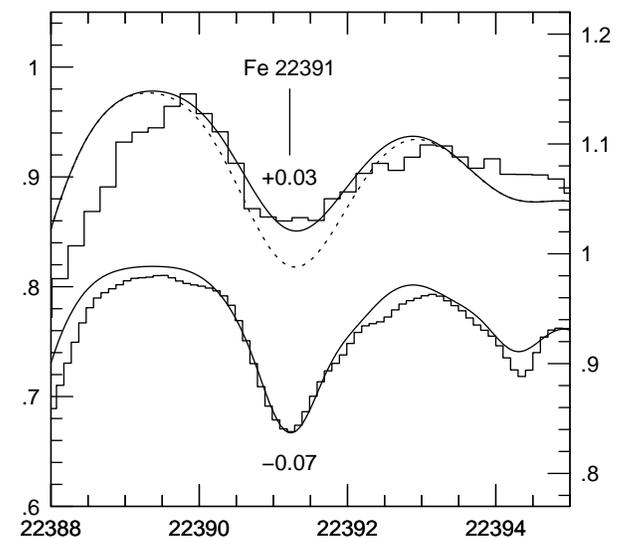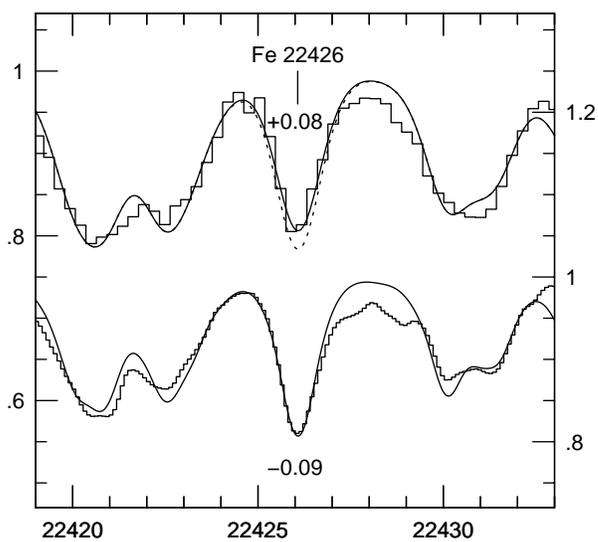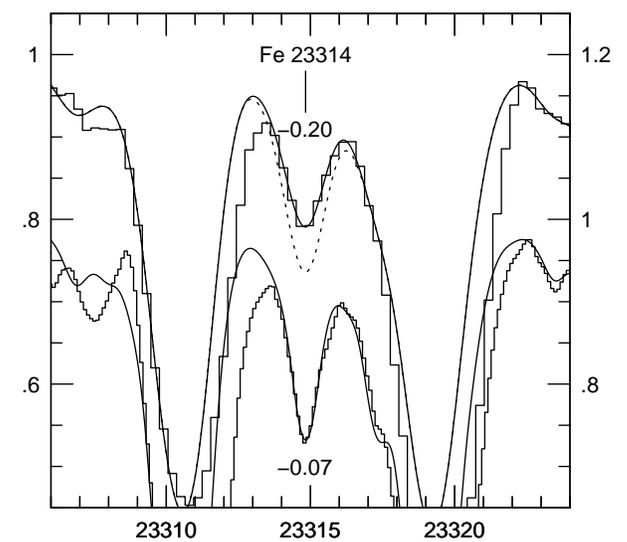

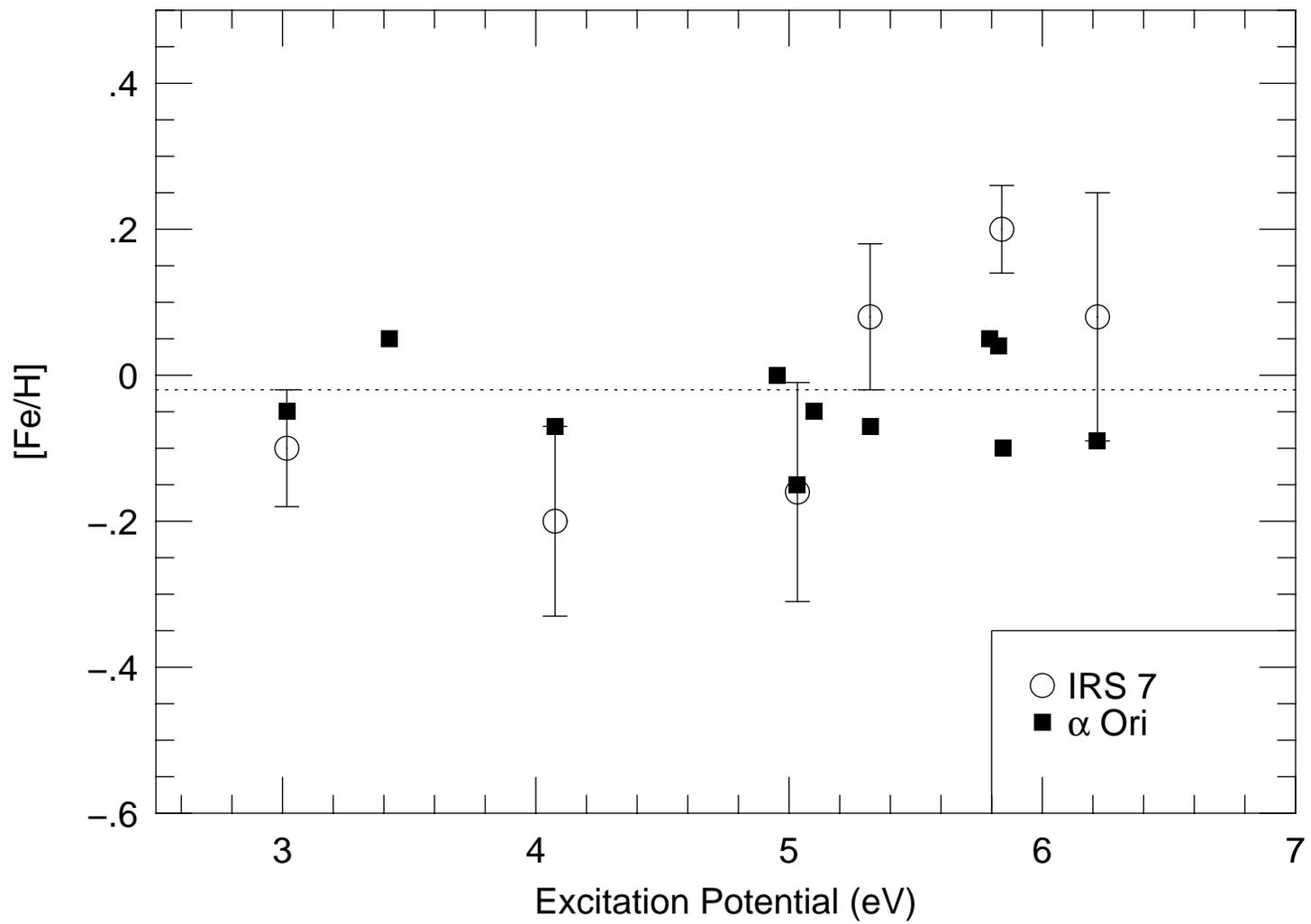

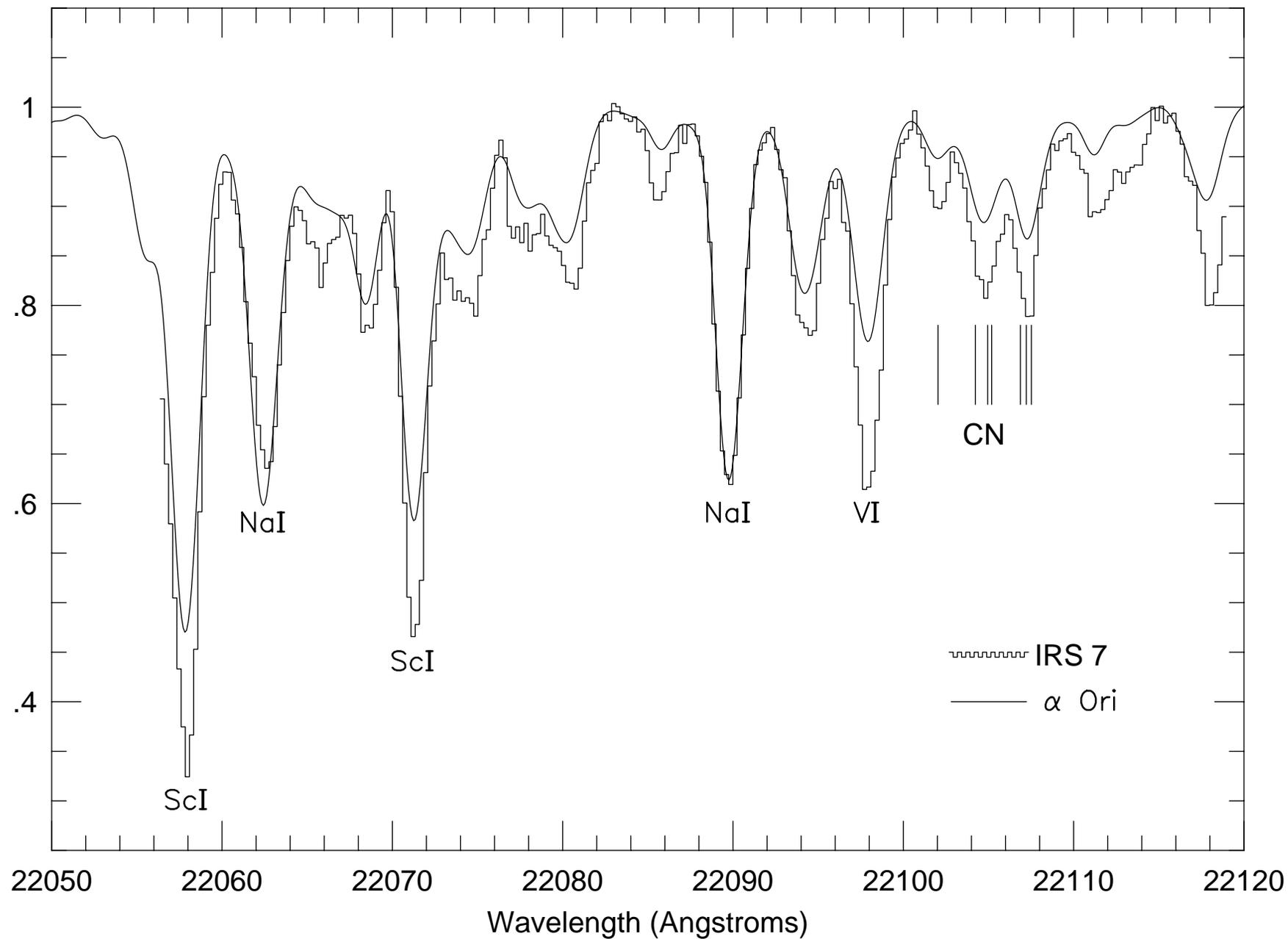

Aboussäid, A., Carleer, M., Hurtmans, D., Biémont, E., & Godefroid, M. R. 1996, Physica Scripta., 53, 28

Abrams, M. C., David. S. P., Rao, M. L. P., Engleman, R., & Brault, J. W. 1994, ApJS, 93, 351

Afflerbach, A., Churchwell, E., & Werner, M. W. 1997, ApJ, 478, 190

Aitken, D. K., Griffiths, J., Jones, B., & Penman, J. M. 1976, MNRAS, 174, 41P

Aitken, D. K., Jones, B., & Penman, J. M. 1974, MNRAS, 169, 35P

Alexander, D. R., Augason, G. C., & Johnson, H. R. 1989, ApJ, 345, 1014

Allen, D. A., Hyland, A. R., & Hillier, D. J. 1990, MNRAS, 244, 706

Balachandran, S. C., & Carney, B. W. 1996, AJ, 111, 946

Becklin, E. E., Matthews, K., Neugebauer, G., & Willner, S. P. 1978, ApJ, 220, 831

Bessel, M. S., & Wood, P. R. 1984, PASP, 96, 247

Black, J. H. 1995, private communication

Blommaert, J. A. D. L., van der Veen, W. E. C. J., Van Langevelde, H. J., Habing, H. J., & Sjouwerman, L. O. 1998, A&A, 329, 991

Blum, R. D., DePoy, D. L., & Sellgren, K. 1995a, ApJ, 441, 603

Blum, R. D., Sellgren, K, & DePoy, D. L. 1995b, ApJL, 440, L17

Blum, R. D., Sellgren, K, & DePoy, D. L. 1996, AJ, 112, 1988

Branch, D., Bonnel, J., & Tomkin, J. 1978, ApJ, 225, 902

Bressan, A., Fagotto, F., Bertelli, G., & Chiosi, C. 1993, A&AS, 100, 647

Brown, J. A., Johnson, H. R., Alexander, D. R., Cutright, L. C., & Sharp, C. M. 1989, ApJS, 71, 623 (BJACS)

Carr, J. S., Sellgren, K., and Balachandran, S. 1996 in The Galactic Center, 4th ESO/CTIO Conference, ASP Conf Ser, 102, 212

Cotera, A. S., Erickson, E. F., Colgan, S. W. J., Simpson, J. P., Allen, D. A., & Burton, M. G. 1996, ApJ, 461, 750

Crowther, P. A., & Dessart, L. 1998, MNRAS, 296, 622

Crowther, P. A., & Smith, L. J. 1997, ApJ, 320, 500

Di Benedetto, G. P. 1993, A&A, 270, 315

Dyck, H. M., Benson, J. A., Ridgway, S. T., & Dixon, D. J. 1992, AJ, 104, 1982

Ertmer, W., & Hofer, B. 1976, Z. Physik A, 276, 9

Figer, D. F., McLean, I. S., & Morris, M. 1995, ApJL, 447, L29

Figer, D. F., McLean, I. S., & Morris, M. 1999, ApJ, 514, 202

Fliegner, J., Langer, N., & Venn, K. A. 1996, A&A, 308, L13

Glass, I. S., Moneti, A., & Moorwood, A. F. M. 1990, MNRAS, 242, 55P

Goorvitch, D. 1994, ApJS, 95, 535

Gratton, R. G., & Sneden, C. 1990, A&A, 234, 366

Gray, D. F. 1977, ApJ, 218, 530

Gray, D. F. 1992, The Observation and Analysis of Stellar Photospheres, 2nd ed. (Cambridge: Cambridge University Press)

Grevesse, N., Noels, A. & Sauval, A. J. 1996, in Cosmic Abundances, ASP Conf Ser, 99, 117

Gustafsson, B., & Jørgensen, U. G. 1994, A&ARev, 6, 19

Habing, H. J., Tigon, J., & Tielens, A. G. G. M. 1994, A&A, 286, 523

Hack, M. Engin, S., Yilmaz, N., Sedmak, G., Rusconi, L., & Boehm, C. 1992, A&AS, 95, 589

Hinkle, K., Wallace, L., & Livingston, W. 1995, PASP, 107, 1042

Holweger, H., Band, A., Kock, M., & Kock, A. 1991, A&A, 249, 545

Holweger, H., & Mueller, E. A. 1974, Sol. Phys., 39, 19

Johnson, H. R., Bernat, A. P., Krupp, B. M. 1980, ApJS, 42, 501

Jørgensen, U. G., & Larsson, M. 1990, A&A, 238, 434

Kleinmann, S. G., & Hall, D. N. B. 1986, ApJS, 62, 501

Krabbe, A., Genzel, R., Drapatz, S., & Rotaciuc, V. 1991, ApJ, 370, 23

Krabbe, A. et al. 1995, ApJ, 447, L95

Kurucz, R. L. 1993, Kurucz CD-ROM No. 13, "ATLAS9 Stellar Atmosphere Programs and the 2 km/s Grid", [Cambridge: Smithsonian Astrophysical Observatory]

Lambert, D. L., Brown, J. A., Hinkle, K. H., & Johnson, H. R. 1984, ApJ, 284, 223

Langer, N. 1992, A&A, 265, L17

Langer, N., Hamann, W.-R., Lennon, M., Najarro, F., Pauldrach, A. W. A., and Puls, J. 1994, A&A, 290, 819

Langer, N., & Heger, A. 1998, in Boulder-Munich II: Properties of Hot, Luminous Stars, ed. I. Howarth, ASP Conf. Series, 131, 76

Lebofsky, M. J., Rieke, G. H., & Tokunaga, A. T. 1982, ApJ, 263, 736

Lee, T. A. 1970, ApJ, 162, 217

Lester, D. F., Bregman, J. D., Witteborn, F. C., Rank, D. M., & Dinerstein, H. L. 1981, ApJ, 248, 524

Lester, D. F., Dinerstein, H. L., Werner, M. W., Watson, D. M., Genzel, R., & Storey, J. W. V. 1987, ApJ, 320, 573

Libonate, S., Pipher, J. L., Forrest, W. J., & Ashby, M. L. N. 1995, ApJ, 439, 202

Livingstone, W., & Wallace, L. 1991, An Atlas of the Solar Spectrum in the Infrared from 1850 to 9000 cm$^{-1}$ (1.1 to 5.4 µm), N.S.O. Technical Report #91-001, (Tucson: National Solar Observatory)

Luck, R. E. 1982, ApJ, 263, 215

McWilliam, A. 1990, ApJS, 74, 1075

McWilliam, A., & Rich, R. M. 1994, ApJS, 91, 749

Maeder, A. 1987, A&A, 178, 159

Maeder, A. 1998, in Boulder-Munich II: Properties of Hot, Luminous Stars, ed. I. Howarth, ASP Conf. Series, 131, 85

Maeder, A., & Conti, P. S. 1994, ARA&A, 32, 227

Martin, P. G., & Whittet, D. C. B. 1990, ApJ, 357, 113

Meynet, G. 1998, in Boulder-Munich II: Properties of Hot, Luminous Stars, ed. I. Howarth, ASP Conf. Series, 131, 96

Meynet, G., Maeder, A., Schaller, G., Schaerer, D., Charbonnel, C. 1994, A&AS ,103 ,97

Moneti, A., Glass, I. S., & Moorwood, A. F. M. 1994, MNRAS, 268, 194

Morris, M. 1993, ApJ, 408, 496

Nagata, T., Woodward, C. E., Shure, M., & Kobayashi, N. 1995, AJ, 109, 1676

Nagata, T., Woodward, C. E., Shure, M., Pipher, J. L., & Okuda, H. 1990, ApJ, 351, 83

Najarro, F., Hillier, D. J., Kudritzki, R. P., Krabbe, A., Genzel, R., Lutz, D., Drapatz, S. & Geballe, T. R. 1994, A&A, 285, 573

Najarro, F., Krabbe, A., Genzel, R., Lutz, D., Kudritzki, R. P., & Hillier, D. J. 1997, A&A, 325, 700

Nave, G., Johansson, S., Learner, R. C. M., Thorne, A. P., & Brault, J. W. 1994, ApJS, 94, 221

Nota, A., Pasquali, A., Drissen, L., Leitherer, C., Robert, C., Moffat, A. F. J., & Schmutz, W. 1996, ApJS, 102, 383

Okuda, H., et al. 1990, ApJ, 351, 890

Pelan, J., & Berrington,K. A. 1995, A&AS, 110, 209

Peterson, R. C., Dalle Ore, C. M., & Kurucz, R. L. 1993, ApJ, 404, 333

Plez, B. 1992, A&AS, 94, 527

Plez, B., Brett, J. M., & Nordlund, A. 1992, A&A, 256, 551

Ramírez, S. V., DePoy, D. L., Frogel, J. A., Sellgren, K., & Blum, R. D. 1997, AJ, 113, 1411

Ramírez, S. V., Sellgren, K., Carr, J. S., Balachandran, S., Blum, R. D. , & Terndrup, D. M. 1999, in The Central Parsecs, Galactic Center Workshop 1998, eds. H. Falcke and A. Cotera, ASP    Conf. Series, in press

Reid, M. J. 1993. ARA&A, 31, 345

Rieke, G. H., & Lebofsky, M. J. 1985, ApJ 288, 618

Rieke, G. H., Rieke, M. J., & Paul, A. E. 1989, ApJ, 336, 752

Ruland, F., Biehl, D., Holweger, H., Griffin, R., & Griffin, R. 1980, A&A, 92, 70

Schaerer, D. 1999, in Active Galactic Nuclei, Dense Stellar Systems, and Galactic Environments, eds. S. Lamb and J. Perry, ASP Conf. Series, in press

Schaerer, D., & Vacca, W. D. 1998, ApJ, 497, 618

Schaller, G., Schaerer, D., Meynet, G., Maeder, A., 1992, A&AS, 96, 269

Sellgren, K., Hall, D. N. B., Kleinmann, S. G., & Scoville, N. Z. 1987, ApJ, 317, 881

Sellmaier, F. H., Yamamoto, T., Pauldrach, A. W. A., & Rubin, R. H. 1996, A&A, 305, L37

Serabyn, E., & Morris, M. 1996, Nature, 382, 602

Serabyn, E., Shupe, D., & Figer, D. F. 1998, Nature, 394, 448

Shields, J. C., & Ferland, G. J. 1994, ApJ, 430, 236

Simpson, J. P., Colgan, S. W. J., Rubin, R. H., Erickson, E. F., & Haas, M. R. 1995, ApJ, 444, 721

Sjouwerman, L. O., Van Langevelde, H. J., Winnberg, A., & Habing, H. J. 1998, A&AS, 128, 35

Smith, V. V., & Lambert, D. L. 1985, ApJ, 294, 326

Smith, V. V., & Lambert, D. L. 1990, ApJS, 72, 387

Sneden, C. 1973, Ph.D. thesis, Univ. of Texas

Stasinska, G., & Schaerer, D. 1997, A&A, 322, 615

Takeda, Y. 1992, A&A, 253, 487

Tamblyn, P., & Rieke, G. H. 1993, ApJ, 414, 573

Tamblyn, P., Rieke, G. H., Hanson, M. M., Close, L. M., McCarthy, D. W., & Rieke, M. J. 1996, ApJ, 456, 206

Terndrup, D. M., Frogel, J. A., & Whitford, A. E. 1991, ApJ, 357, 453

Tokunaga, A. T., Toomey, D. W., Carr, J. S., Hall, D.N.B. & Epps, H. W. 1990, SPIE, 1235, 131

Tolansky, S. 1948, Hyperfine Structure in Line Spectra and Nuclear Spin, 2nd ed. (London: Methuen)

Tsuji, T. 1991, A&A, 245, 203

Wallace, L., & Hinkle, K. 1996, ApJS, 107, 312

Willner, S. P., Russell, R. W., Puetter, R. C., Soifer, B. T., & Harvey, P. M. 1979, ApJ, 229, L65

Wood, P. R., Habing, H. J., & McGregor, P. J. 1998, A&A, 336, 925

Zeiske, W., Meisel, G., Gebauer, H., Hofer, B., & Ertmer, W. 1976, Phys Lett, 55A, 405

TABLE 1

CSHELL Spectra

| Central Wavelength (Å) | Features Used | Signal-to-noise Ratio | |
|---|---|---|---|
| | | IRS 7 | VV Cep |
| 16297 | weak CO; OH | 50 | 270 |
| 21202 | Fe I | 90 | 370 |
| 22086 | Na doublet | 90 | 300 |
| 22393[a] | Fe I | 135 | … |
| 22416 | Fe I | 140 | 280 |
| 23120 | moderate CO | 125 | 350 |
| 23299 | Fe I | 110 | … |
| 23402 | CO 2-0 R(99) | 140 | … |

[a] Obtained with the $256^2$ SBRC array; all other spectra obtained with the $256^2$ NICMOS array.

TABLE 2

CO Line Data and Equivalent Widths for Giants

| Transition[a] | λ (Å) | χ (eV) | log gf | Equivalent Width (mÅ) | |
|---|---|---|---|---|---|
| | | | | α Boo | β And |
| 6-3 R(10) | 16266.75 | 0.813 | -6.765 | 42.8 ± 1.0 | … |
| 6-3 R(57) | 16271.20 | 1.546 | -5.792 | 46.4 ± 1.5 | 97 ± 3 |
| 6-3 R(58) | 16278.37 | 1.572 | -5.779 | 41.6 ± 1.1 | 93 ± 2 |
| 6-3 R(8) | 16280.68 | 0.804 | -6.862 | 34.0 ± 0.9 | 83 ± 3 |
| 6-3 R(59) | 16285.84 | 1.599 | -5.766 | 37.1 ± 1.2 | 73 ± 3 |
| 6-3 R(60) | 16293.63 | 1.626 | -5.753 | 34.8 ± 3.0 | … |
| 6-3 R(6) | 16295.77 | 0.797 | -6.981 | 28.4 ± 1.7 | 81 ± 2 |
| 5-2 P(13) | 16300.06 | 0.571 | -7.139 | 33.9 ± 1.8 | 86 ± 3 |
| 5-2 R(81) | 16310.97 | 2.049 | -5.820 | 8.6 ± 0.9 | … |
| 6-3 R(4) | 16312.01 | 0.792 | -7.136 | 16.4 ± 2.1 | 56 ± 9 |
| 5-2 P(14) | 16313.69 | 0.577 | -7.111 | 36.7 ± 1.1 | 95 ± 2 |
| 6-3 R(63) | 16318.86 | 1.710 | -5.715 | 29.0 ± 2.7 | 59 ± 2 |
| 2-0 R(79) | 23103.40 | 1.476 | -4.915 | 207.0 ± 2.5 | … |
| 2-0 R(80) | 23115.68 | 1.513 | -4.906 | 192.0 ± 3.0 | 289 ± 7 |
| 2-0 R(81) | 23128.41 | 1.550 | -4.898 | 186.0 ± 2.0 | 275 ± 2 |
| 2-0 R(82) | 23141.59 | 1.587 | -4.890 | 187.0 ± 2.0 | … |
| 2-0 R(83) | 23155.23 | 1.625 | -4.881 | 171.0 ± 1.8 | 282 ± 8 |
| 2-0 R(92) | 23298.93 | 1.984 | -4.810 | 102.6 ± 2.4 | 168 ± 4 |
| 2-0 R(93) | 23317.27 | 2.026 | -4.802 | 97.2 ± 1.0 | 172 ± 8 |
| 2-0 R(94) | 23336.10 | 2.069 | -4.795 | 92.6 ± 0.9 | 160 ± 8 |
| 2-0 R(99) | 23437.66 | 2.286 | -4.757 | 91.0 ± 1.2 | … |

[a] Wavelengths, lower excitation potential, and gf-values are from Goorvitch (1994).

TABLE 3

CO Equivalent Widths (mÅ) for Supergiants

| Transition[a] | IRS 7 | VV Cep | α Ori |
|---|---|---|---|
| 6-3 R(58) | … | … | 263 ± 19 |
| 6-3 R(8) | … | … | 263 ± 11 |
| 6-3 R(60) | … | 253 ± 10 | … |
| 5-2 P(13) | 196 ± 12 | 294 ± 10 | 273 ± 10 |
| 6-3 R(4) | 110 ± 16 | 167 ± 10 | 179 ± 14 |
| 6-3 R(63) | 111 ± 16 | 164 ± 10 | 180 ± 12 |
| 2-0 R(80) | 439 ± 17 | 532 ± 23 | 524 ± 13 |
| 2-0 R(81) | 456 ± 18 | 531 ± 19 | 526 ± 6 |
| 2-0 R(83) | 424 ± 13 | 519 ± 16 | 498 ± 17 |
| 2-0 R(99) | 170 ± 16 | … | 254 ± 10 |

[a] Wavelengths, lower excitation potential, and gf-values are listed in Table 2.

TABLE 4

Dependence of $T_{eff}$, $\xi$ and C Abundance on Model Atmosphere

| Model  | β And            | α Ori            | VV Cep           |
|--------|------------------|------------------|------------------|
| Kurucz | 3830, 2.00, 8.32 | 3580, 3.14, 8.31 | 3590, 3.53, 8.24 |
| BJACS  | 3800, 2.74, 8.26 | 3530, 3.83, 8.21 | 3540, 4.40, 8.14 |
| Plez   | 3800, 2.16, 8.31 | 3540, 3.23, 8.27 | 3480, 3.67, 8.18 |

Entries for each star are the derived effective temperature in K, microturbulence in km s$^{-1}$, and carbon abundance, using the respective model atmosphere.

TABLE 5

Fe I Line Data and Equivalent Widths for Sun and α Boo

| Wavelength[a] (Å) | $\chi$[a] (eV) | log gf[b] | Equivalent Width (mÅ) | |
|---|---|---|---|---|
| | | | Sun | α Boo |
| 21183.98 | 3.017 | -4.187 | 10.8 ± 0.5 | 89.0 ± 2.0 |
| 21229.95 | 6.764 | -0.727 | 10.0 ± 1.5 | 15.4 ± 1.0 |
| 21244.30 | 4.955 | -1.277 | 84.0 ± 2.0 | 158.4 ± 3.0 |
| 21781.16 | 5.874 | -2.229 | 2.0 ± 0.3 | 5.0 ± 0.6 |
| 21781.82 | 3.415 | -4.425 | 2.7 ± 0.3 | 28.0 ± 1.5 |
| 21793.75 | 6.400 | -0.841 | 16.0 ± 2.0 | 26.1 ± 2.5 |
| 21819.96 | 5.848 | -1.583 | 9.0 ± 0.7 | 22.3 ± 1.0 |
| 21838.98 | 6.780 | -0.343 | 22.6 ± 1.0 | 25.0 ± 2.5 |
| 22046.29 | 6.532 | -1.089 | 7.0 ± 0.5 | 13.0 ± 1.5 |
| 22085.89 | 5.849 | -1.602 | 9.0 ± 0.5 | 17.3 ± 4.0 |
| 22350.55 | 5.720 | -1.625 | 11.0 ± 0.5 | 20.7 ± 1.4 |
| 22381.29 | 5.844 | -1.458 | 12.0 ± 0.6 | 20.9 ± 1.0 |
| 22386.94 | 5.033 | -0.481 | 229.0 ± 8.0 | 281.0 ± 12.0 |
| 22391.22 | 5.320 | -1.600 | 25.0 ± 2.0 | 71.0 ± 2.0 |
| 22399.02 | 5.099 | -1.249 | 72.0 ± 3.0 | 136.0 ± 3.0 |
| 22426.08 | 6.218 | -0.224 | 79.0 ± 3.0 | 91.0 ± 1.5 |
| 22818.82 | 5.792 | -1.244 | 22.0 ± 2.0 | 51.0 ± 6.0 |
| 22838.60 | 5.099 | -1.336 | 66.0 ± 1.0 | 127.0 ± 3.0 |
| 22852.17 | 5.828 | -0.652 | 66.0 ± 2.0 | 125.0 ± 3.0 |
| 23314.83 | 4.076 | -2.665 | 32.0 ± 2.0 | 125.0 ± 17.0 |

[a] Wavelength and lower excitation potential from Nave et al. (1994).

[b] Semi-empirical solar gf-values.

TABLE 6

Fe I Equivalent Widths (mÅ) for M Giants

| Wavelength[a] | β And | 30 Her | HR 6702 |
|---|---|---|---|
| 21183.98 | 230 ± 10 | … | … |
| 21244.30 | 290 ± 5 | … | … |
| 21781.16 | 13 ± 2 | … | … |
| 21781.82 | 94 ± 2 | 146 ± 6 | 130 ± 12 |
| 21819.96 | 46 ± 2 | 78 ± 3 | 104 ± 6 |
| 22085.89 | 37 ± 3 | 66 ± 4 | 79 ± 4 |
| 22350.55 | 51 ± 3 | … | … |
| 22381.29 | … | 80 ± 5 | 79 ± 4 |
| 22386.94 | 372 ± 7 | 465 ± 18 | 493 ± 10 |
| 22391.22 | 124 ± 4 | 181 ± 8 | 179 ± 4 |
| 22426.08 | … | 203 ± 10 | 191 ± 6 |
| 22818.82 | 99 ± 7 | 162 ± 3 | 154 ± 6 |
| 22838.60 | 240 ± 10 | … | … |
| 23314.83 | 231 ± 20 | … | … |

[a] Lower excitation potential and gf-values are listed in Table 5.

TABLE 7

Fe I Abundances[a] for M Supergiants

| Wavelength[b] | α Ori | VV Cep | IRS 7 |
|---|---|---|---|
| 21183.98 | -0.05 ± 0.06 | -0.08 ± 0.09 | -0.15 ± 0.08 |
| 21244.30 | 0.00 ± 0.07 | … | … |
| 21781.82 | +0.05 ± 0.03 | … | … |
| 22381.29 | -0.10 ± 0.03 | … | +0.20 ± 0.06 |
| 22386.94 | −0.15 ± 0.05 | -0.26 ± 0.15 | -0.16[c] ± 0.15 |
| 22391.22 | −0.07 ± 0.04 | -0.05 ± 0.05 | +0.08[c] ± 0.10 |
| 22426.08 | -0.09 ± 0.07 | +0.15 ± 0.16 | +0.08 ± 0.17 |
| 22818.82 | +0.05 ± 0.03 | … | … |
| 22838.60 | -0.05 ± 0.05 | … | … |
| 22852.17 | +0.04 ± 0.05 | … | … |
| 23314.83 | -0.07 ± 0.04 | … | -0.20 ± 0.13 |

[a] [Fe/H] from spectrum synthesis.

[b] Lower excitation potential and gf-values are listed in Table 5.

[c] Average value from two spectra.

TABLE 8

Uncertainty in [Fe/H] for IRS 7

| | |
|---|---|
| $T_{eff} \pm 230$ K | $\mp 0.04$ |
| $\xi \pm 0.30$ km s$^{-1}$ | $\mp 0.08$ |
| log g $\pm 0.2$ | $\pm 0.06$ |
| standard error | $\pm 0.07$ |
| Total uncertainty | $\pm 0.13$ |

TABLE 9

Results for [Fe/H]

| Star | Number Lines | Model Grid | | |
| --- | --- | --- | --- | --- |
| | | Kurucz | BJACS | Plez |
| α Boo | 20 | -0.49 ± 0.02[a] | … | … |
| β And | 12 | 0.04 ± 0.09 | 0.00 ± 0.08 | -0.03 ± 0.08 |
| 30 Her | 8 | … | … | 0.00 ± 0.11 |
| HR6702 | 8 | … | … | +0.16 ± 0.11 |
| α Ori | 11 | 0.00 ± 0.08 | -0.02 ± 0.08 | -0.04 ± 0.08 |
| VV Cep | 4 | -0.03 ± 0.12 | -0.07 ± 0.14 | -0.06 ± 0.13 |
| IRS 7 | 6 | … | … | -0.02 ± 0.13 |

[a] Uncertainty is only the standard error in the mean for α Boo; all other stars also include the uncertainties due to effective temperature, microturbulence and gravity.